\documentclass[aps,pre,twocolumn,superscriptaddress,showkeys,showpacs,longbibliography]{revtex4-2}
 \usepackage[version=4]{mhchem}
\usepackage{amsmath,amssymb,graphicx,bm}
\usepackage{tikz}
\usepackage{siunitx}
\usepackage{comment}
\usepackage[normalem]{ulem} % for \sout
\usepackage{hyperref}
\usepackage{bm}
\usepackage{graphicx}

\definecolor{klccolor}{RGB}{0, 102, 204}   % blue for changes
\definecolor{klcolor}{RGB}{200, 0, 0}    % red for comments
\definecolor{klxcolor}{gray}{0.5}         % gray for strikeout

\newcommand{\dg}{{^\dagger}}
\newcommand{\bk}{{\bf k}}
\newcommand{\br}{{\bf r}}

\newcommand{\pmat}[1]{\begin{pmatrix} #1 \end{pmatrix}}
\newcommand{\bs}{\bm{\sigma}}
\newcommand{\sbk}{\sin \bk\!{\raisebox{-0.8ex}{$\scriptstyle\perp$}}\kern-5pt\cdot\!\bs}
\begin{document}

% \title{Emergent Spin Triplet Superconducting State from Local Parity-Breaking hybridization in \ce{CeRh2As2}}
%\title{Local parity breaking hybridization induced even and odd pairing in \ce{CeRh2As2}}
\title{Kondo Lattice model for field induced triplet pairing in CeRh$_2$As$_2$}
\author{Indra Gankhuyag}
\affiliation{Center for Materials Theory, Rutgers University, Piscataway, NJ 08855, U.S.A.}
\author{Piers Coleman}
\affiliation{Center for Materials Theory, Rutgers University, Piscataway, NJ 08855, U.S.A.}
\affiliation{Hubbard Theory Consortium, Department of Physics, Royal Holloway, University of London,
Egham, Surrey TW20 0EX, United Kingdom.}
\date{\today}

\begin{abstract}
Spin triplet superconductors are of interest because of their topological properties, and their robustness to high magnetic fields. CeRh$_2$As$_2$ is a unique heavy fermion superconductor with both
even-parity and candidate odd-parity superconducting phases with a special feature enabling a field induced phase transition between the two. The field induced transition which switches the parity of the Cooper pairs is enabled by a staggered antisymmetric spin orbit coupling (ASOC) interaction.  However, the heavy fermion origins of the superconductivity in CeRh$_2$As$_2$ and the origin of the antisymmetric spin-orbit coupling are currently unexplored. Here, we develop a Kondo lattice model for CeRh$_2$As$_2$ and  demonstrate how local hybridization between Cerium $d$- and $f$-orbitals,  driven by the locally broken inversion symmetry,  at the Cerium sites generates the ASOC. The resulting Kondo lattice model leads to nearly degenerate even and odd parity superconducting states that display a field induced parity transition between an even-parity state and an odd-parity pseudospin triplet state.  We discuss the wider implications of our results extended to other heavy fermion systems like CeSiI. 
\end{abstract}
\maketitle
\section{Introduction}
The possibility of superconductivity that survives strong magnetic fields has renewed interest in odd-parity and spin-triplet pairing, not least because some such states may possess nontrivial topological properties. In spin orbit coupled materials, however, the physical spin of a Cooper pair need not coincide with the pseudospin character of the gap projected onto the low energy quasiparticle bands. This distinction is especially important in multiband heavy fermion superconductors, where crystal symmetry and strong electronic correlations act together to determine the pairing state.

The heavy fermion compound \ce{CeRh2As2} exhibits a field-driven transition between two superconducting phases~\cite{cerh2as2,cerh2as2_local_moment}. For a magnetic field along the crystallographic $c$ axis, the low field SC1 phase gives way to a high field SC2 phase near $4\,\mathrm{T}$, while superconductivity persists to an upper critical field of approximately $14\,\mathrm{T}$, far above the conventional weak coupling Pauli scale for $T_c\simeq0.3\,\mathrm{K}$. The pronounced contrast in field response motivates the interpretation of SC1 and SC2 as competing even and odd parity states, although the microscopic spin structure of the high field phase remains an active question~\cite{cerh2as2_coexistence_field}.

\begin{figure}[ht]
    \centering
    %\hspace*{-0.2\columnwidth}
    \includegraphics[width=1.0\columnwidth]{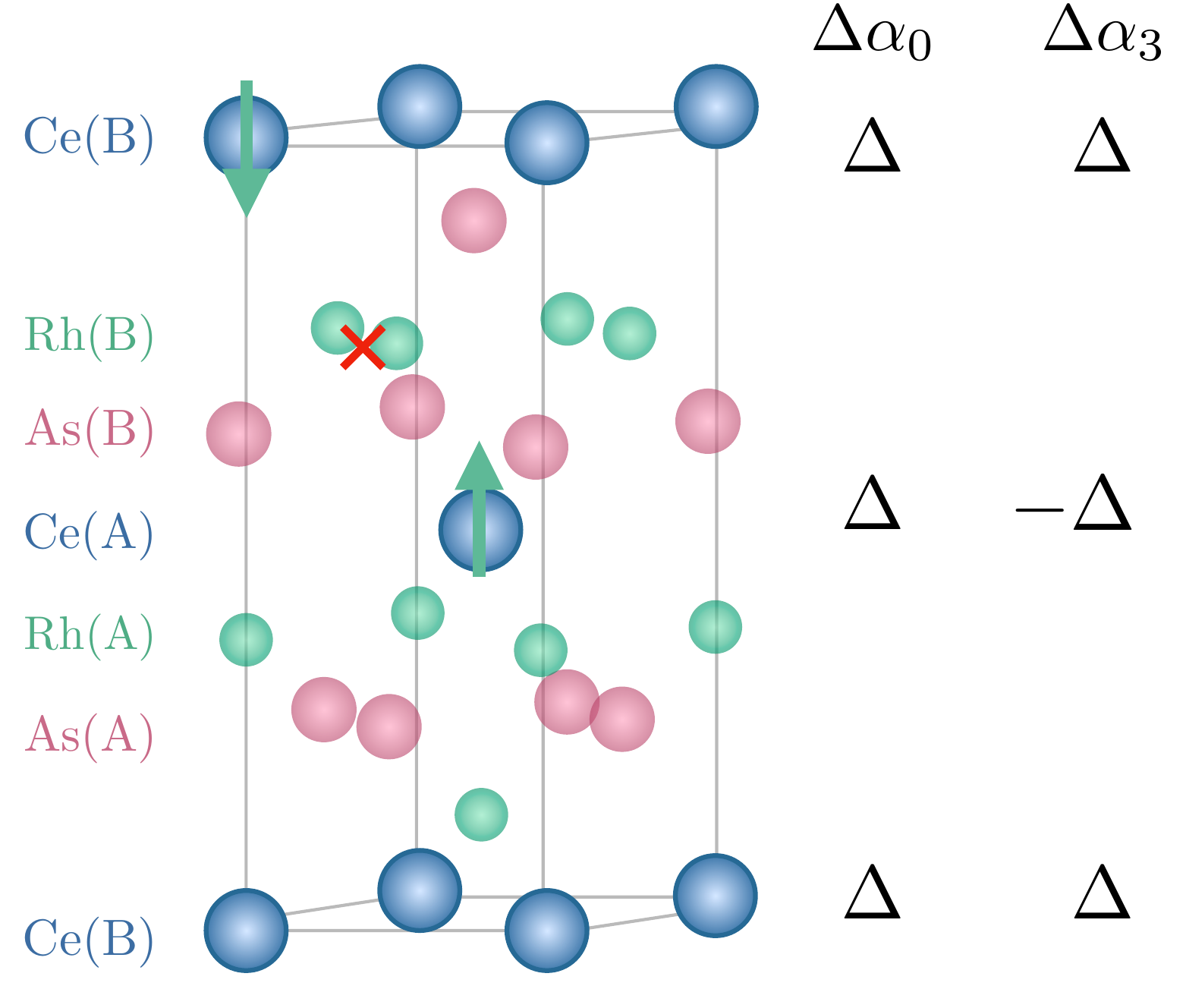}
     \caption{Left: the  crystal structure of \ce{CeRh2As2} which crystallises in a tetragonal $P4nmm$ structure. The inversion center is represented by a red cross in between two distinct \ce{Ce} atoms leading to a unit cell with two atoms. These are the Ce(A) and Ce(B) which are related to one another by inversion. There is the electric potential coming from the Rh atoms depicted by green arrows. Right: the two pairing possibilities stabilized by staggered ASOC interaction. The $\Delta\alpha_0$ is the uniform solution which has the same sign pairing gap and $\Delta\alpha_3$ is the staggered solution which has a modulated pairing gap on alternating layers. Here, $\Delta$ is the pairing gap magnitude and $\alpha_m$ is the sublattice degree of freedom where $m=0,3$ which are the Pauli matrices.
   % \klc{Add another subfigure which shows the lattice geometry and possibility hopping parameters used for the normal state.}
   }
    \label{fig:cerh2as2_lattice}
\end{figure}

A key feature of \ce{CeRh2As2} is that its crystal is globally inversion symmetric but the two Ce sites within a unit cell each lack a local inversion center. Inversion exchanges the Ce sublattices, allowing a staggered antisymmetric spin orbit coupling (ASOC) whose sign alternates between the layers. Symmetry based descriptions have shown how this structure can stabilize uniform even-parity and staggered odd parity superconducting states and allow a magnetic field to favor the latter~\cite{nonsymmorphic,pdw,parity_cerh2as2}. The corresponding pairing patterns are illustrated in Fig.~\ref{fig:cerh2as2_lattice}.

Several theoretical approaches have clarified the competition between superconducting states in locally noncentrosymmetric systems. M{\"o}ckli and Ramires~\cite{MockliRamires2021Scenarios} analyzed both even to odd and odd to odd field driven transitions within a two sublattice description, while Cavanagh \textit{et al.}~\cite{nonsymmorphic} emphasized the role of nonsymmorphic symmetry and the associated momentum dependent spin orbit structure. Nogaki and Yanase~\cite{NogakiYanase2022Parity} demonstrated that electronic correlations can enhance the parity transition field. Extending the description to vortex degrees of freedom, Minamide and Yanase~\cite{MinamideYanase2025Meron,MinamideYanase2026Vortices} investigated the parity transition and possible unconventional vortex textures.

Recent microscopic studies provide complementary perspectives on the pairing mechanism. Lee, Agterberg, and Brydon~\cite{LeeAgterbergBrydon2025} found competing superconducting instabilities of opposite parity and spin density wave tendencies associated with type-II van Hove singularities. Nogaki and Yanase~\cite{NogakiYanase2026Geometry}, using a multiorbital Dirac--Anderson model, identified leading $B_{1u}$ and $B_{2g}$ superconducting instabilities influenced by quantum geometry and correlations. The appearance of a $B_{1u}$ channel provides a useful symmetry level comparison with the leading projected gap discussed below, although the microscopic origins and orbital structures of the states need not coincide. Work on the coexistence of magnetic order and superconductivity~\cite{AminWuShishidouAgterberg2024} further underscores the importance of the nearby ordered phases.

More recently, Ishizuka \textit{et al.}~\cite{Ishizuka2024} investigated the superconducting symmetry and possible topological properties of \ce{CeRh2As2} using realistic electronic structure calculations and crystalline symmetry analysis. In particular, their classification accommodates staggered spin singlet pairing in an odd parity representation, including a $d_{x^2-y^2}$-type form factor in the $B_u$ sector appropriate to a field along the $c$ axis. This provides a useful symmetry benchmark for a microscopic theory. The symmetry classification alone, however, does not determine how the staggered ASOC is generated in a heavy fermion Kondo lattice, or how the corresponding odd parity gap develops on the heavy quasiparticle Fermi surface.

These questions are sharpened by the strongly correlated normal state. The Curie--Weiss susceptibility, the onset of coherent transport on the Kondo scale, and the large superconducting specific heat anomaly indicate the importance of local moments and heavy quasiparticles~\cite{cerh2as2_local_moment,cerh2as2,cerh2as2_pressure}. Neutron measurements also reveal pronounced quasi-two-dimensional antiferromagnetic fluctuations~\cite{Broholm24}, which naturally motivate predominantly intralayer $d$-wave spin singlet pairing. The challenge is to understand how such pairing can develop in a Kondo lattice with antiferromagnetic interactions, while also giving rise to an odd parity pseudospin triplet gap in the high field phase.

Here we construct a minimal Kondo lattice description in which local inversion symmetry breaking permits an additional onsite hybridization channel. Interference between this channel and the conventional intersite hybridization generates a staggered ASOC for the heavy quasiparticles. We then study the competition between uniform and staggered intralayer pairing states in a magnetic field, determining their phase boundary by comparing the self-consistent free energies. The staggered physical spin singlet order parameter acquires an effective odd parity pseudospin triplet structure upon projection onto the heavy bands. Its leading momentum dependence is compatible with the $B_u$ symmetry class discussed by Ishizuka \textit{et al.}, while our calculation identifies a microscopic route to that structure.

The paper is organized as follows. Section~II introduces the crystal symmetries and the uniform and staggered pairing channels. Section~III develops the local parity breaking hybridization and its symmetry allowed form factors. Section~IV constructs the Kondo lattice model and mean-field formulation. Section~V presents the emergent ASOC, the field dependent superconducting phase diagram, and the projected pseudospin gap. Section~VI discusses the physical interpretation, limitations, and possible extensions to related heavy fermion materials. Technical details of the decoupling and charge conjugation conventions are given in the appendices.

\section{Crystal Symmetry and Pairing}
CeRh$_2$As$_2$ has a tetragonal $P4/nmm$ space group with two Ce atoms per unit cell. Although there is global inversion symmetry, locally, inversion symmetry is broken at the Ce sites due to the alternating RhAs and AsRh layers along the $c$ axis. The two Ce atoms in the unit cell are related by inversion, as shown in Fig.~\ref{fig:cerh2as2_lattice}. This leads to an additional sublattice degree of freedom and allows a staggered antisymmetric spin orbit coupling (ASOC) interaction of the form:
\begin{equation}
H_{\rm ASOC}=\Lambda \alpha_3(\sin k_y \sigma_1 - \sin k_x \sigma_2),
\end{equation}
where $\alpha_i$ and $\sigma_i$ are Pauli matrices in sublattice and spin space respectively. The sublattice $\alpha_3$ reverses sign under sublattice exchange and together with the momentum inversion, preserves global inversion. 

In conventional superconductors, the allowed pairing symmetry is determined from the combined parity of spin and momentum which must be antisymmetric. The electron pair wavefunction with fermionic statistics must be antisymmetric upon particle exchange:
\begin{equation}\label{exchange_fermion}
     \Delta(1,2)=-\Delta(2,1)
\end{equation}
where $1=\{ \mathbf{R}_1,\alpha_1,\sigma_1\}$ and
$2=\{\mathbf{R}_2,\alpha_2,\sigma_2\}$ represents the position, sublattice, and spin of particle 1 and 2 respectively. Typically, in conventional superconductors with inversion symmetry, the allowed states are momentum even spin singlet or momentum odd spin triplet states.
\begin{table*}[t]
\centering
\caption{The symmetry allowed pairing states in \ce{CeRh2As2}. The total fermionic pair wavefunction must have odd parity upon particle exchange. The pairing wavefunction is given by $\Delta_{ s}$, where $s=\rm A,B$ which represent pairing within the layers which represent pairing between the same sublattice \ce{Ce} atoms. The total parity is obtained by the product of the parities which include spin $\mathcal{P}_\sigma$, momentum $\mathcal{P}_l$, and sublattice $\mathcal{P}_\alpha$ parity. 
% The in-plane singlet even in momentum pairing can either have an even or odd parity superconducting state depending on the how the sublattice transforms.
}
\label{tab:antisymmetry}
\renewcommand{\arraystretch}{1.3}
\begin{tabular}{c c c c c c c c}
\hline\hline
Spin & $\mathcal{P}_\sigma$ & Momentum & $\mathcal{P}_l$ &  Sublattice &  $\mathcal{P}_\alpha$ & $\mathcal{P}$ & Total Parity \\
\hline
Singlet & $(-)$ &  $\,f_{\mathbf{k}} = f_{-\mathbf{k}} \,$&$(+)$ & $\Delta_{\rm A}=\Delta_{\rm B}$ & $(+)$ & $(+)$ &
Even \\
 Singlet & $(-)$ &  $\,f_{\mathbf{k}} = f_{-\mathbf{k}} \,$&$(+)$ &  $\Delta_{\rm A}=-\Delta_{\rm B}$ & $(-)$ & $(-)$ & Odd \\
 Triplet & $(+)$ & $\,\vec{d}_{\mathbf{k}} = -\vec{d}_{-\mathbf{k}}$ & $(-)$& $\Delta_{\rm A}=\Delta_{\rm B}$ & $(+)$ & $(-)$ & Odd \\
 Triplet & $(+)$ & $\,\vec{d}_{\mathbf{k}} =- \vec{d}_{-\mathbf{k}}$ & $(-)$ & $\Delta_{\rm A}=-\Delta_{\rm B}$ & $(-)$ & $(+)$ & Even \\
\hline\hline
\end{tabular}
\end{table*}
However, \ce{CeRh2As2} has an additional sublattice degree of freedom that enables a new combination of spin, momentum and sublattice. The quasi two dimensional nature of \ce{CeRh2As2} restricts to intralayer dominant pairing within the Ce layers. The symmetry allowed pairing states are given in Table.\ref{tab:antisymmetry}.
% \begin{align}
% \Delta_{AA} &= \phantom{-}\Delta_{BB},
% \\
% \Delta_{AA} &= -\Delta_{BB}.
% \end{align}
% where $\Delta_{AA}$ represents pairing within the layers. 

The two pairing solutions allowed by the sublattice is the uniform state which has the same pairing sign on all layers corresponding to an even parity state and the staggered state which has different pairing sign on alternating layers corresponding to an odd parity state. The two distinct solutions are shown in Fig. \ref{fig:cerh2as2_lattice}. These two states are nearly degenerate in energy and exhibit a field induced transition in the presence of a field applied perpendicular to the plane.
% On the Fermi surface, the even and odd parity states gives singlet and triplet pairing respectively. 
% In the limit of strong ASOC and weak interlayer coupling, two states are nearly degenerate in energy and exhibits a field induced transition in the presence of a field applied perpendicular to the plane.
% These staggered ASOC has enabled two nearly degenerate pairing states with distinct parity.
% The staggered state has an odd form factor from the sublattice structure which leads to a triplet state on the Fermi surface despite the underlying pairing is spin singlet.
% \begin{equation}
% \Delta_{\alpha\beta}^{\sigma\sigma'}(\mathbf{k})=
% -
% \Delta_{\beta\alpha}^{\sigma'\sigma}(-\mathbf{k}).
% \end{equation}
% \begin{align}
% \Delta\alpha_0&: (\Delta,\Delta), \\
% \Delta\alpha_3 &: (\Delta,-\Delta).
% \end{align}
\section{Parity breaking onsite hybridization}
 \begin{figure}[ht]
    \centering
    \includegraphics[width=0.9\columnwidth]{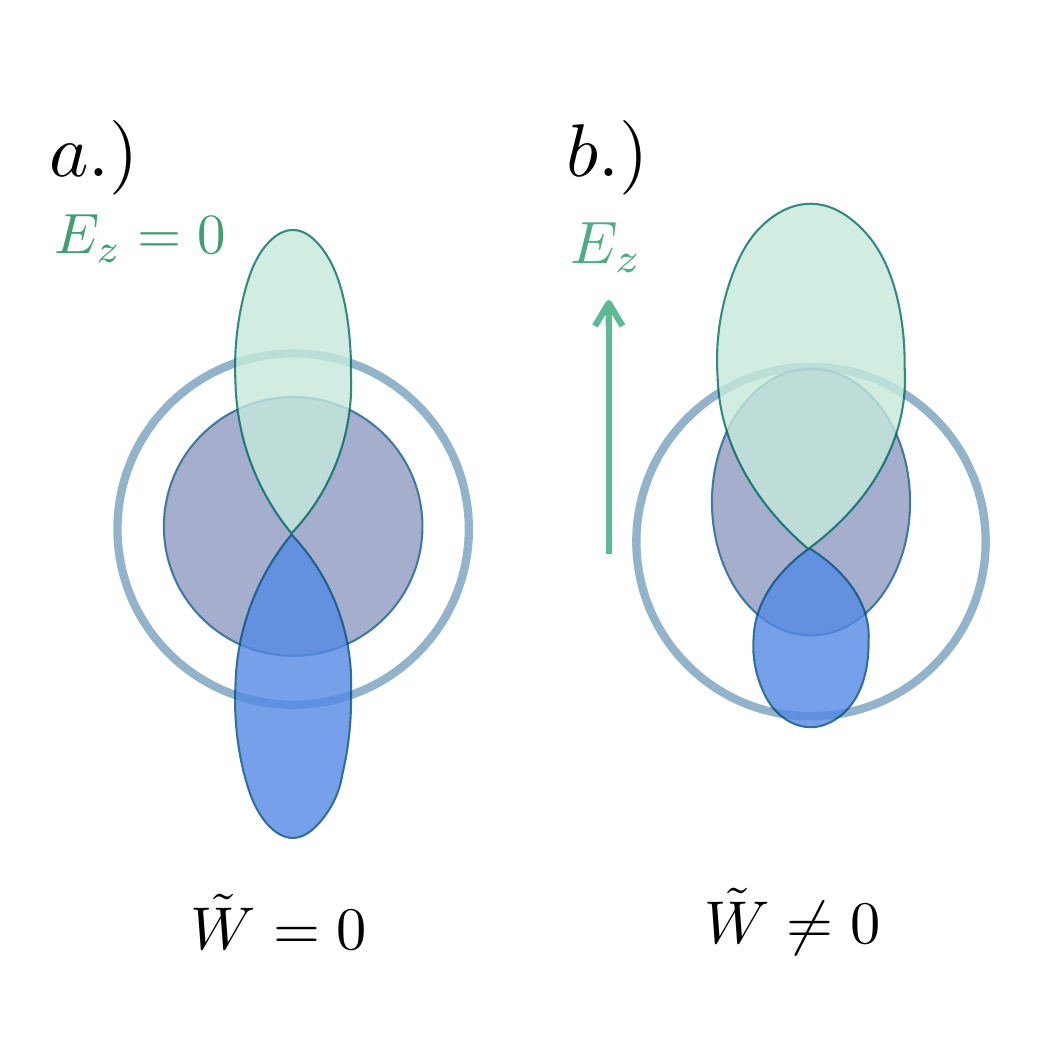}
    \caption{The depiction of orbitals overlapping on an atom. Here $E_z$ is the electric field along the $z$ direction and $\tilde W$ is hybridization strength. In green and blue is the $p$-orbital which has a node and in purple is the isotropic $s$-orbital.  a.) Inversion center is located at the atomic center which has an even atomic potential $V(x)=V(-x)$. Orbitals have well defined parity. The spatial overlap integral between $p$ and $s$ orbitals is zero, $\tilde W=0$ due to the atomic potential is even. b.) Inversion is broken at the atomic sites, the local crystal fields distort the orbitals leading to an offset of the inversion center. The atomic potential is no longer even $V(x)\neq V(-x)$. The resulting overlap integral between $p$ and $s$ orbitals is finite $\tilde W\neq0$.}\label{fig:onsite_hybridization}
\end{figure}
% This section will address the origin of the staggered ASOC interaction enabled by a new local hybridization mechanism. 
A simple model of  Ce based heavy fermion systems treats the high energy physics via an Anderson model, describing $f$ electrons hybridized with a conduction band formed from delocalized $d$ electrons. The low energy  physics is then obtained by carrying out a Schrieffer-Wolff transformation that encodes the physics of the hybridization as a Kondo exchange interaction between the local moments and the conduction sea. If we also include the nearest neighbor RKKY interactions, the resulting
Hamiltonian is called a Kondo Heisenberg lattice. 

The hybridization between an $f$-state  and a $d$-state is given by the overlap of the wavefunctions with the atomic potentials.  The local hybridization is then
\begin{equation}\label{overlap}
    V_{fd}=\int_{-\infty}^\infty d^3\mathbf{r}\psi^*_{f\sigma}(\mathbf{r})V(\br)\psi_{d\sigma'}(\br),
\end{equation}
where $V(\mathbf{r})$ is the atomic potential at a Cerium site. In locally centrosymmetric systems, the $f$-state is odd parity $\psi_{f\sigma}({\bf r}) = - \psi_{f\sigma}({- \bf r})$ while the $d$-state is even parity $\psi_{d\sigma}({\bf r}) = + \psi_{d\sigma}({- \bf r})$, and since the electronic potential is inversion symmetric, $V(\mathbf{r})= V(-\mathbf{r})$,
$V_{fd}=0$ vanishes.  
% \begin{equation}\label{even_pot}
%     V(x)=V(-x).
% \end{equation}
%Then the overlap integral between $f$ and $d$ electrons given by Eq.\ref{overlap} is odd under inversion resulting to a vanishing local hybridization.
Thus in a centrosymmetric crystal, the $f-d$ hybridization takes place between neighboring Cerium sites and when written in momentum space, must be odd under inversion, yet even under time reversal. A minimal model for the non local odd parity hybridization within the planes of Ce atoms, has the form
\begin{equation}\label{nonlocal}
 V\sbk = V (\sin k_x \sigma_x + \sin k_y \sigma _y),
\end{equation}
% $ V_0\sin(\mathbf{k})\cdot \vec{\sigma}$
where $V$ is the intersite hybridization strength and $\bk_{\perp}$ is the in-plane wavevector. This model form is obtained in a Slater-Koster treatment  of hybridizing $s$- and $p$-states and captures the essence of spin orbit coupled heavy fermion compounds allowing, for example, a simple description of topological Kondo insulators \cite{topology_Kondo}.
% A typical odd parity hybridization found in centrosymmetric systems is given by
By contrast, \ce{CeRh2As2} has a locally noncentrosymmetric structure with inversion centers located in between the Cerium layers. In this case, the atomic potential contains a gradient along the $c$-axis that alternates between neighboring Cerium layers. This time, the atomic potential is not necessarily inversion symmetric:
\begin{equation}
     V_\alpha(x,y,z)= V_{-\alpha}(-x,-y,-z) \neq V_{\alpha}(-x,-y,-z),
\end{equation}
where $\alpha=\pm 1$ is a layer index.  In this situation, the 
the $f$- and $d$- states now acquire a $c$-axis dipole moment which alternates between layers, and they no longer have a well defined parity. 
The resulting overlap between the $f$ and $d$ states at the same site given by \eqref{overlap} is then non-zero, leading to a local hybridization. A minimal model for the local parity breaking hybridization  is given by:
% $ iW\alpha_3\sigma_3$,
\begin{equation}\label{local}
    i\tilde W\alpha_3\sigma_3,
\end{equation}
where $\tilde W$ is the local parity breaking hybridization amplitude.
The appearance of the $\alpha_3$ term reflects the reversal of the $c$-axis electric field on neighboring layers.  The hybridization is invariant under inversion, which inverts both the spin and sublattice index. The imaginary prefactor also guarantees that the hybridization is invariant under time-reversal. 
A model calculation  of such a local hybridization mechanism is obtained by calculating the Slater-Koster overlap integral between a $j=1/2$ $s$ and $p$-state (see: Fig.\ref{fig:onsite_hybridization}). A more realistic model of the hybridization, involving the $j=5/2$ overlap between a $d$ and $f$ state is expected to give rise to an even momentum dependence in the quantity $W$ which we neglect here.  
Combining the local \eqref{local} and intersite \eqref{nonlocal} terms, we can formulate a minimal model for the hybridization form factor
\begin{equation}\label{onsite_intersite}
    \phi_{\mathbf{k},\sigma\sigma'}=({\bf s}_{\bk}\cdot \bm{\sigma}+i{W}\alpha_3\sigma_3).
\end{equation}
where ${W}=\tilde {W}/V$ and ${\bf s}_{\bk} = (\sin k_x,\sin k_y)$. The dimensionless parameter $W$ is the ratio of the local hybridization amplitude $\tilde W$ to the intersite amplitude $V$. For real $W$ and $V$, the form factors obey the exact identity
\begin{equation}\label{phi_identity}
\phi_{\bk}\phi_{\mathbf{k}}\dg=(s_x^2+s_y^2+W^2)\alpha_0\sigma_0+2W\alpha_3(s_y\sigma_1-s_x\sigma_2).
\end{equation}
The interference term is staggered in the layer index and has the antisymmetric spin-orbit structure.
%The first term is the intersite odd parity hybridization and the second term is the new local parity breaking hybridization. 
When an $f$ electron tunnels through the conduction band, the resulting tunneling amplitude will reflect an interference between the two terms in the hybridization. We can calculate this effect by integrating out the conduction electrons, to obtain an $f$ electron self-energy given by 
\begin{equation}
    \Sigma_f({\bf k},i\omega_n) \sim  \frac{V^2 }{i\omega_n - \epsilon_{\bk} }\phi_{\bk}\phi_{\mathbf{k}}\dg
\end{equation}
where for simplicity we have approximated the conduction electrons by a single band of dispersion $\epsilon_{\bf k}$.
The interference between the two channels creates a spin orbit cross term in this self-energy that has the form
\begin{equation}
    \Sigma_f(\mathbf{k})\propto VW\alpha_3(s_y\sigma_1-s_x\sigma_2).
\end{equation}
where we denote $s_x=\sin k_x, \ s_y = \sin k_y$. 
This spin orbit term has the same momentum and spin dependence as the previous staggered ASOC interaction introduced from symmetry analysis. The Kondo lattice therefore demonstrates a mechanism of how the ASOC is generated through a local parity breaking hybridization induced  by  the locally broken inversion symmetry.

\section{Model and Methods}

\subsection{Kondo lattice construction}

We now formulate a minimal Kondo Heisenberg model for \ce{CeRh2As2},  
with the Hamiltonian $H={H}_c+{H}_K+{H}_H$ where
\begin{eqnarray}\label{conduction}
{H}_c&=&\sum_{\mathbf{k}}c\dg_{\mathbf{k},\alpha,\sigma}(h_{\mathbf{k}})_{\alpha\alpha'}c_{\mathbf{k},\alpha',\sigma},
\\ \label{kondo}
    {H}_K&=&J_K\sum_j{\tilde c}^\dagger_{j,\sigma}\vec{\sigma}_{\sigma\sigma'}{\tilde c}_{j,\sigma'} \cdot {\mathbf{S}}_{j},    
\\ 
\label{heisenberg}
   {H}_H&=&J_H\sum_{\langle i,j\rangle}{\mathbf{S}}_{i}\cdot{\mathbf{S}}_{j},
\end{eqnarray}
where $c_{\bk,\alpha,\sigma}^\dagger$ creates a conduction electron with momentum $\bk$, sublattice index $\alpha=\pm 1$ and spin index $\sigma = \pm 1$, while ${\mathbf{S}}_j= (S_j^x,S_j^y,S_j^z)$ is the three component spin-$1/2$ local moment operator for site $j$. We use Pauli matrix normalization for the local moments, i.e ${\rm Tr}[S_j^{\sigma}S_j^{\sigma'}]=2 \delta^{\sigma \sigma'}$. A summation convention is used for the sum over spin and sublattice indices $\alpha$. 
% For simplicity, the operator hats for the creation and annihilator operators are dropped as ollows. $\hat c_{j,\sigma}\dg=c_{j,\sigma}\dg$ and $\hat c_{j,\sigma}=c_{j,\sigma}$
The first term $H_c$ describes the conduction electrons, with a dispersion obtained from a tight binding calculation on a tetragonal lattice
\begin{equation}
h_{\mathbf{k}}=\epsilon_{\mathbf{k},00}\alpha_0+\epsilon_{\mathbf{k},10}\alpha_1.
\end{equation}
Here the dispersion elements  
\begin{equation}
    \begin{aligned}
      & \epsilon_{\mathbf{k},00}=-2t(\cos(k_x)+\cos(k_y))-\mu, \\
      & \epsilon_{\mathbf{k},10}=-4t_s\cos\left(\frac{k_x}{2}\right)\cos\left(\frac{k_y}{2}\right)\cos\left(\frac{k_z}{2}\right) \\
    \end{aligned}
\end{equation}
describe the intra, and inter-layer hopping processes. The normal state parameters are given by $t$ the intralayer hopping strength and $\mu$ the chemical potential and $t_s=t_u+t_d$ is the interlayer hopping strength, $t_u$ is hopping up and $t_d$ is hopping down.
To maintain a connectivity with earlier work, we have followed the
notation used in \cite{cerh2as2,nonsymmorphic}. 

$H_K$ is the local Kondo interaction between the conduction electrons and the local moments, where $J_K$ is the Kondo coupling strength and the site index $j\equiv ({\bf R}_j,\alpha) $ is a  proxy for the position of the unit cell and the sublattice index of the Cerium site.  Following the discussion in the introduction, the  operator $\tilde{c}_{j,\sigma}$ annihilates a conduction electron in a Wannier state localized at site $j$:
\begin{equation}
    \tilde{c}_{j,\alpha,\sigma}=\frac{1}{\sqrt{\mathcal{N}_s}}\sum_{\mathbf{k}}
\phi(\mathbf{k})_{\alpha,\sigma;\alpha',\sigma'}c_{\mathbf{k},\alpha',\sigma'} e^{i\mathbf{k}\cdot\mathbf{R}_j},
\end{equation}
where $\mathcal{N}_s$ is the number of unit-cells in the crystal. The form factor matrix $\phi_{\mathbf{k}}=\phi({\mathbf{k}})$ describes a Wannier state which carries the symmetry of the hybridization inherited from an underlying Anderson model.  We take this to have the sublattice diagonal form introduced in \eqref{onsite_intersite},
\begin{equation}\label{formfactor}
    \phi_{\mathbf{k}}=({\bf s}_{\bk}\cdot \bm{\sigma}+i{W}\alpha_3\sigma_3).
\end{equation}
where we use the short-hand ${\bf s}_{\bk} = (\sin k_x,\sin k_y)$.
The dominant source of spin orbit coupling in a Kondo lattice derives from the hybridization with the $f$  electrons. In our minimal model spin orbit coupling only enters through these terms in the Kondo coupling, and for simplicity, all spin orbit coupling terms in the conduction sea have been dropped. 

Lastly, the Heisenberg interaction in Eq.\ref{heisenberg} describes the AFM superexchange interaction between neighbouring spins where $J_H$ is the coupling strength. 
% The Kondo Heisenberg model describes the interplay between the development of magnetic ordering favoured by the Heisenberg interaction and the heavy Fermi liquid from the Kondo interaction. 

We employ a  large-$N$ path integral approach to  formulate the model as a field theory and then develop a mean-field description. In the large-$N$ limit, all quantities that are extensive in $N$ behave as semi-classical variables, whose fluctuations are suppressed by powers of $1/N$, causing the path integral to focus into a mean field configuration. In the limit $N\rightarrow \infty$, the mean field solution becomes exact, allowing the basis for a controlled mean-field expansion. 

An important aspect of our work is to use a large $N$ generalization of the SU($2$) group to the symplectic group Sp($N$), where $N$ is even \cite{Read91,flint_natphys2008,flintSymplecticTimeReversal2009}. The first member of the Sp$(N)$ family {\sl is} the  SU$(2)$ group, $\hbox{Sp}(2)=\hbox{SU}(2)$, but at larger $N$, Sp$(N)\in\hbox{SU}(N)$ is a subgroup of SU$(N)$ that preserves  time reversal symmetry. Sp($N$) loosely corresponds to an $N/2$-fold replication of the up and down spins, corresponding to $N/2$ ``up'' and $N/2$ ``down'' spin components:  
\begin{equation}
    \sigma \in \begin{pmatrix} \uparrow_1, & \uparrow_2\dots \uparrow_{N/2}\cr
    \downarrow_1, & \downarrow_2\dots \downarrow_{N/2}
        \end{pmatrix}\equiv  (\pm 1, \pm 2 \dots \pm N/2 )
\end{equation}
where we identify the spin labels $\uparrow_{\sigma}\equiv \sigma$ and $\downarrow_{\sigma} \equiv -\sigma$, where $\sigma = (1, 2, \dots N/2) $ label the replicas \cite{flintSymplecticTimeReversal2009}. The spin operators that connect up and down spins in the form factor \eqref{formfactor} are diagonal in replica space which allows us to replicate the spins at each site to controlling the mean field theory without interfering with the SU($2$) spin algebra which rotates between up and down spins in the original model. Our final results are expressed in terms of the properties per spin pair component, equivalent to formally setting $N=2$ at the end of the calculation. 

To formulate our model as a field theory, we use a fermionic representation 
\begin{equation}
    (\mathbf{S}_j)^{\lambda} = f\dg_{j\alpha}S^{\lambda}_{\alpha\beta}f_{j\beta},\qquad \lambda \in {\textstyle (1, \frac{N(N+1)}{2})} , 
\end{equation}
for  the components of the Hermitian Sp$(N)$ spin operators, where $f\dg_{\alpha}$ and $f_{\beta}$ are fermionic creation and annihilation operators with spin component $\alpha \in (\pm 1, \dots \pm N/2)$. In passing to the large-$N$ theory we make the conventional rescaling $J_H\rightarrow J_H/N$ and  $J_K\rightarrow J_K/N$, so that the interaction energy remains extensive in $N$ and the $N\rightarrow\infty$ limit is well defined. 
Using  the corresponding Fierz decomposition (see appendix \ref{app:spn_decoupling} for details), we rewrite the Kondo and Heisenberg interactions in the form: 
\begin{eqnarray}
    {\tilde c}^\dagger_{j,\sigma}\vec{\sigma}_{\sigma\sigma'}{\tilde c}_{j,\sigma'} \cdot {\mathbf{S}}_j&\rightarrow& (\tilde c\dg_{j,\sigma}\tilde c_{j,\sigma'}) {\rm S}_{\sigma'\sigma}(j)\cr
    {\mathbf{S}}_{i}\cdot{\mathbf{S}}_{j}&\rightarrow&{\rm S}_{\sigma\sigma'}(i){\rm S}_{\sigma'\sigma}(j)
\end{eqnarray}
where the
\begin{equation}
    {\rm S}_{\sigma\sigma'}(j) = f^\dg_{j,\sigma} f_{j,\sigma'} -
    \tilde{\sigma}\tilde{\sigma}'f^\dg_{j,-\sigma} f_{j,-\sigma'},\quad (\tilde{\sigma}\equiv\text{sgn}(\sigma))
\end{equation}
describe the matrix unit form of the  spin operators.  (These corresponds to an antisymmetrization of their  SU$(N)$  spin counterpart ${\rm S}^{{\rm SU}(N)}=f^\dg_{\alpha} f_{\beta} - (n_f/N)\delta_{\alpha\beta} $ under time reversal). 
% The second term is the particle hole transformation of the first term $f_\sigma\rightarrow \tilde{\sigma}f^\dagger_{-\sigma}$ which ensures particle-hole symmetry.
% The first and second term in Eq.\ref{kondo_spn} and Eq.\ref{heisenberg_spn} represents the particle-hole and particle-particle channels respectively. 
The parton decomposition of the local moments gives rise to a local SU($2$) gauge invariance 
\begin{equation}\label{gauge_invariance}
    f_{j,\sigma}\rightarrow \left\{
    \begin{array}{c}
         e^{i\phi_j}f_{j,\sigma} , \cr
         \cos{\theta_j} f_{j,\sigma} + \sin{\theta_j}\tilde \sigma f^\dagger_{j,-\sigma},
    \end{array}
    \right.
\end{equation}
which corresponds to rotations in particle-hole, or isospin space. 
We select amongst the various possible representations of the local moment by requiring that the $f$ parton state is an isospin singlet, requiring
\begin{equation}\label{Gutzwiller}
\begin{aligned}
 & \sum_{\sigma>0}f^\dagger_{i,\sigma}f^\dagger_{i,-\sigma}= \sum_{\sigma>0}f_{i,-\sigma}f_{i,\sigma}=0,\,\cr
 & n_f = N/2 \\
\end{aligned}      
\end{equation}
corresponding to a half-filled $f$ state with local no $s$-wave pairing. These are the ``Gutzwiller constraints'' of the Kondo lattice, 
which inside a path integral, are naturally imposed by fluctuations over the SU(2) gauge fields. 
%The first two constraints prohibit local onsite pairing and the last constraint ensures single occupancy at each site requiring $ n_{f,\sigma}=N/2$, for $N$ spin components. Formally, the large-$N$ expansion for Sp($N$) is constructed for $N$ spin components. In this problem, the number of spin components will be set to $N=1$ representing $S=1/2$ case.
% $\langle \vec{W}_j\cdot\vec{\tau}\rangle=0$ where $\vec{W}_j=(W_{1},W_{2},\lambda_j)$ are the strength of the constraint fields and $\vec{\tau}$ are particle-hole Pauli matrices. The constrains $W_{1}=W_1(j)$ and $W_{2}= W_{2}(j)$ constraints p. The last constraint $\lambda_j$ conserves particle number at each site requiring $n_f=1$ where $n_f=f^\dagger_{j,\alpha,\sigma}f_{j,\alpha,\sigma}$.  
% The path integral is described in terms of coherent states and the quartic interactions are required to be factorised into quadratic form to go inside the path integral. 

\subsection{Path integral formulation}
The SU($2$) path integral is given by the partition function as $Z=\text{Tr}[e^{-\beta H}P_{G}]$ where $P_{G}$ imposes the Gutzwiller constraints  \eqref{Gutzwiller}. To develop the path integral, we adopt a Balian Werthammer notation, 
\begin{equation}
    C_{{\bf k},\alpha,\sigma}= \pmat{
        c_{{\bf k},\alpha, \sigma}\cr c_{{\bf k},\alpha, -\sigma}\cr
        - c\dg_{-{\bf k},\alpha, 
        - \sigma}\cr c\dg_{-{\bf k},\alpha,  \sigma}}, \ 
    F_{j,\sigma }=\pmat{
        f_{j,\sigma}\cr f_{j,-\sigma}\cr -f\dg_{j,-\sigma}\cr f\dg_{j,\sigma}
    },  (\sigma > 0),
        \end{equation}
where the spin labels  are purely positive.  We will 
 combine these Nambu spinors into a row, as follows
\begin{eqnarray}
    { C}_{\bk,\alpha }&=& \bigl({ C}_{\bk, \alpha 1}, { C}_{\bk ,\alpha 2}, \dots { C}_{\bk,\alpha, N/2}\bigr)\cr
       { F}_{j }&=& \bigl({ F}_{j,  1}, { F}_{j,2}, \dots { F}_{j,N/2}\bigr).
\end{eqnarray}
In terms of these fields, the Gutzwiller constraints become ${ F}_j\dg\vec \tau { F}_j=0$ where $\vec \tau = (\tau_1,\tau_2,\tau_3)$ are the isospin Pauli matrices, acting in particle-hole space.
We can also form the conduction  Wannier states as
\begin{equation}
    { C}_{j,\alpha} = \frac{1}{\sqrt{\mathcal{N}_s}}\sum_{\mathbf{k}, \alpha'}
\phi_{\mathbf{k},\alpha\alpha'} {C}_{\mathbf{k},\alpha'}e^{i\mathbf{k}\cdot\mathbf{R}_j}
\end{equation}
where the invariance of the form factor \eqref{formfactor} under charge conjugation  allows us to extend its usage to the  Nambu formulation of the Wannier states. (See appendix \ref{app:charge_conjugation} for details). 
In terms of this representation, the Gutzwiller constraint is simply that the isospin vanishes 
${ F}\dg_{j}\vec \tau_j{ F}_{j}=0$, where the $\vec{\tau}$ are the Pauli matrices in the particle-hole space of the Nambu spinors.
In terms of this representation, the path integral has the form:
\begin{equation}\label{path}
Z=\int D[\bar{\psi},\psi,w]\text{exp}\Big[-{\int^\beta_0 d\tau(\mathcal{L}_0+H)_{}}\Big],
\end{equation}
where the Berry phase term 
\begin{equation}
    \mathcal{L}_0= \sum_{\bk,\alpha,\sigma}
{C}^\dagger_{\bk,\alpha,\sigma}\partial_\tau{C}_{\bk, \alpha ,\sigma}+\sum_{j,\sigma}{F}_{j,\sigma}^\dagger(\partial_\tau+ w_j){F}_{j,\sigma}.
\end{equation} 
Here  $w_j=\vec w_j\cdot \vec\tau$, where $\vec{w}_j=(w^1_{j},w^2_{j},\lambda_j)$ are three SU(2) gauge potentials whose  fluctuation  along the imaginary axis impose the Gutzwiller constraints (\ref{Gutzwiller}). 
%The last constraint $\lambda_j$ ensures \sout{single} 
%{\color{red} fixed} occupancy %{\color{red}$n_f=N/2$} at each site, where $n_{j\sigma}=f^\dagger_{j\sigma}f_{j\sigma}$. 

 \begin{widetext}
The Fierz identities of  the Sp($N$) representation allow the Heisenberg and Kondo interactions to be rewritten in the form (see Appendix \ref{app:spn_decoupling} for details)
\begin{eqnarray}
\label{heisenberg_spn}
H_H &=&
 -\frac{J_H}{N}\sum_{\langle i,j\rangle}\big[f^\dagger_{j,\sigma}f_{i,\sigma}f^\dagger_{i,\sigma'}f_{j,\sigma'}
+\tilde{\sigma}\tilde{\sigma}'f^\dagger_{i,\sigma}f^\dagger_{j,-\sigma}f_{j,-\sigma'}f_{i,\sigma'}\big] \\
\label{kondo_spn}
H_K &=&
-\frac{J_K}{N}\sum_j\big[\tilde{c}^\dagger_{j,\sigma}f_{j,\sigma}f^\dagger_{j,\sigma'}\tilde{c}_{j,\sigma'}
+\tilde{\sigma}\tilde{\sigma}'\tilde{c}^\dagger_{j,\sigma}f^\dagger_{j,-\sigma}f_{j,-\sigma'}\tilde{c}_{j,\sigma'}\big] ,
 \end{eqnarray}
where the spin sums run over $\sigma, \sigma' \in (\pm 1, \dots \pm N/2)$. %There are prefactors from the symplectic decoupling absorbed into the coupling constant as $J_H=J_H/4$ and $J_K=J_K/2$.  
 \end{widetext}
These interactions are now decoupled using the Hubbard Stratonovich transformations. The Kondo interaction becomes 
\begin{equation}\label{kondo_decoupled}
H_K\rightarrow\sum_{j}\left[
    {C
}^\dagger_{j}{\cal V}\dg_j
 {F}_{j}+ 
 {F
}^\dagger_{j}{\cal V}_j
 {C}_{j}
+N\frac{\text{Tr}[{\cal V}_j\dg{\cal V}_j]}{2J_K} \right],
\end{equation}
where the matrix
\begin{equation}\label{hyb}
  {\cal V}_j= \begin{pmatrix}
        V_j & \nu_{j}^*\\
        \nu_{j} &-V_j^*
    \end{pmatrix}.
\end{equation}
describes hybridization ($V_j$) and pairing ($\nu_j$) between the conduction and $f$ parton fields. 
Similarly, a Hubbard Stratonovich  decoupling of the Heisenberg term gives 
%{}
\begin{equation}\label{heisenberg_decoupled}
H_H\rightarrow\sum_{\langle i,j\rangle}\left[
{F}^\dagger_{i}
 U_{ij} {F}_{j}+ {F}^\dagger_{j}U_{ij}^\dagger
 % Indra please do not reverse the order of the ij in the above equation. 
     {F}_{i}
+N\frac{{\rm Tr}[U^\dagger_{ij} U_{ij}]}{2J_H}\right].
\end{equation}
where
\begin{equation}
    U_{ij}=  \begin{pmatrix}
        t_{ij}^f & \Delta_{ij}^*\\
        \Delta_{ij} &- t_{ij}^{f*}
    \end{pmatrix}, U_{ji}= \pmat{t_{ij}^{f*} & \Delta_{ij}^*\cr
        \Delta_{ij} &- t_{ij}^{f}}.
\end{equation}
Here, the $t_{ij}^f$ and $\Delta_{ij}$ are the $f$ electron hopping and RVB pairing fields, respectively. Notice also that $U_{ji}=U_{ij}^\dagger$. The notation $\langle i,j\rangle$ denotes nearest neighbours within a given layer. Note that the nearest neighbour pairing $\Delta_{ij}\equiv \Delta_{{\bf R}_i,\alpha;{\bf R}_j, \alpha}$ can depend on the layers.  

With these transformations, the Lagrangian in the path integral now takes the form 
\begin{equation}
    {\cal L}_0 = \sum_{\bk,\alpha} { C}\dg_{\bk,\alpha}\partial_{\tau}C_{\bk,\alpha} +  \sum _j { F}^\dagger_j(\partial _{\tau} + w_j) F_j +H.
\end{equation}
Here $H=H_c+H_K+H_H$, while $H_K$ take the expanded forms shown in \eqref{kondo_decoupled} and \eqref{heisenberg_decoupled}, respectively. 
The SU$(2)$ gauge invariance \eqref{gauge_invariance} now includes the gauge fields, 
 \begin{eqnarray}
     { F}_j \rightarrow g_j { F}_j,  \quad
    \left\{ \begin{array}{rcl}
 {\cal V}_j&\rightarrow& g_j{\cal V}_j,\cr
        w_j & \rightarrow& g_jw_jg_j^\dagger+ g_j \partial_{\tau }g_j^\dagger , \cr
         U_{ij} &\rightarrow & g_iU_{ij}g_j^\dagger.
     \end{array}  \right.
 \end{eqnarray}
     where $g_j $ are members of the SU$(2)$ group. We use this gauge degree of freedom to fix the gauge by writing ${\cal V}_j= g_j |V_j| \tau_3$,  absorbing the $g_j$ by the above  gauge transformation. In this ``radial gauge", ${\cal V}_j = |V_j| \tau_3 $ describes a pure, real hybridization.    
The pure Kondo model, without the Heisenberg terms describes a  Fermi liquid with an expanded Fermi surface volume which accommodates the hybridized $f$ electrons, according to Oshikawa's theorem\cite{oshikawa,hazra21}. In this phase, the difference between the internal SU$(2)$ gauge field and the external electromagnetic field is Higgsed\cite{Coleman}, in a fashion reminiscent of electroweak symmetry breaking. The finite mass of the difference field locks the internal gauge field to the external electromagnetic field, endowing the previously neutral $f$ partons with physical charge, thus converting them into heavy electrons. Once the Heisenberg coupling is restored, there are now further low temperature instabilities of the heavy Fermi liquid.   

\subsection{Mean field Theory}
To develop a mean field theory  we examine static, translationally invariant solutions.  In the normal state,
$V_j=V$ and $\lambda_j=\lambda$, and  $t_{ij}^f=-t_f$ are real, while the pairing order parameter variables $\Delta_{ij}=0$ are zero.  We restrict our study of instabilities of the Fermi liquid to superconductivity, in which the nearest neighbour Heisenberg couplings develop an expectation value that can depend on the sublattice variable $\alpha$. Within a layer, energetics favours a $d$-wave  pairing state, i.e $\Delta_x = - \Delta_y = \Delta$. The $d$-wave order parameter within the layers guarantees there is no $s$-wave pairing at the $f$ sites, which allows us to set the pairing components of the constraint fields to zero, i.e $\vec w = (w^1,w^2,\lambda)= (0,0,\lambda)$.

Now since the inversion centers lie between the  layers, this allows for the possibility of both uniform and staggered  pairing solutions in which the pairing develops a layer dependence.  These two classes of solution are respectively, even and odd-parity under inversion.
\begin{eqnarray}
    \Delta_{\bk}^{(\alpha)} = 2 \Delta (\cos k_x- \cos k_y) \times\left\{ \begin{array}{cl}
         1 & \hbox{even parity }  \\
        \alpha &  \hbox{odd-parity }
    \end{array}\right.
\end{eqnarray}
where $\alpha = \pm 1$ is the sublattice label. We will show that under the influence of the hybridization, these two states  these correspond to singlet and triplet pairing states, as originally reasoned by Yoshida, Sigrist and Yanase\cite{parity_mixed_noncentrosymmetric}. To accomodate a magnetic field, we introduce a Zeeman term into the Hamiltonian ${\cal H}\rightarrow {\cal H} - \sigma_z B$. We will show that while the uniform state has a higher $T_c$, the staggered odd parity triplet state is able to accomodate a Zeeman polarization, stabilizing it in a  magnetic field.  

Putting this all together, the resulting mean-field Hamiltonian 
takes the form
\begin{eqnarray}\label{mean_field_H}
H_{\rm MF}
= \sum_{\bk}\Psi_{\bk}\kern-3pt\dg
  {\cal H}(\bk)\Psi_{\bk}
+2N{\cal N}_s
\scalebox{0.8}{$
  \displaystyle
  \left[
    \frac{|V|^2}{J_K}
    +\frac{2(|\Delta|^2+t_f^2)}{J_H}
  \right]
$},
\end{eqnarray}
where the superconducting Hamiltonian is given by 
\begin{equation}\label{scham}
     {\cal H}(\bk) = \pmat { 
    h_{\bk}\tau_3 -\sigma_z B& {\cal V}_{\bk}\dg
    \cr 
    {\cal V}_{\bk} & h^f_\bk\tau_3 -\sigma_z B + \Delta^{(\alpha)}_{\bk }\tau_1}
\end{equation}
and
\begin{eqnarray}
    \Psi_{\bk} = \pmat{{ C}_{\bk +}\cr { C}_{\bk -}\cr { F}_{\bk +}\cr { F}_{\bk -}}, %\, \Psi\dg_{\bk} = ({ C}\dg_{\bk +},{ C}\dg_{\bk -}, { F}\dg_{\bk +}, { F}\dg_{\bk -}),
\end{eqnarray}
combines the conduction and $f$ Nambu spinors on the two layers. Here $\alpha =\pm $ are the layer indices of ${ C}_{\bk,\alpha}$ and ${ F}_{\bk ,\alpha} $. The off diagonal hybridization terms are given by ${\cal V}_{\bk} = V \phi_\bk\tau_3$.  The $f$ electron dispersion is
\begin{equation}
    h^{f}_\mathbf{k}=-2t_f(\cos(k_x)+\cos(k_y))+\lambda.
\end{equation}  
In a compact notation define $h^f_\bk=t_f\gamma_\bk+\lambda$, where $\gamma_{\bk}=-2(\cos(k_x)+\cos(k_y))$. Note that the factor of $2N{\cal N}_s$ in the mean-field term derives from the $N$ spin components and the $2{\cal N}_s$ sites.

The mean field solutions are obtained by finding the stationary points of the free energy. Integrating out the fermionic fields leads to the free energy $F=-T\sum_{\mathbf{k},n}\ln(\det(-\mathcal{G}_\mathbf{k}^{-1}(i\omega_n))$ where $\mathcal{G}_\mathbf{k}^{-1}(i\omega_n)=[i\omega_n-\cal {H}(\mathbf{k})]$ is the Greens function in imaginary time where $i\omega_n\beta=i\pi(2n+1)$ represents the fermionic Matsubara frequencies that runs over $n$ discrete frequencies. The mean field Hamiltonian is diagonalised by integrating out the $f$ electrons. With the energy eigenvalues $\xi_{\mathbf{k}\eta}$, using the identity $\ln(\det(A))=\text{Tr}[\ln(A)]$ and computing the Matsubara sum leads to the mean field free energy
\begin{align}\label{free_energy}
\frac{F}{N}=
&\frac{1}{2}\sum_{\mathbf{k}, \eta=1,8}\left[-2T\ln\left(2\cosh\left(\frac{\xi_{\mathbf{k}\eta}}{2 T}\right)\right)\right]\\
&+2{\cal N}_s\left(\frac{|V|^2}{J_K}+\frac{2(t_f^2+|\Delta|^2)}{J_H}\right),
\end{align}
where  $\pm \xi_{\bk n}$ are the eigenvalues of the mean-field Hamiltonian, $\eta\in(1,8)$ is the eigenvalue index and ${\cal N}_s$  the number of unit cells in the lattice \footnote{We note that the original Hamiltonian is 16 dimensional, however, particle-hole  reduces the number of positive eigenvalues to 8. In the absence of a magnetic field, there is also a Kramer's degeneracy, leading to 4 distinct eigenvalues. The pre-factor of $1/2$ derives from the use of a Balian Werthammer four component representation, which requires summation over just half the Brillouin zone. }.     The saddle point approximation is obtained by seeking the stationary points of the effective action ${\delta S_{\rm eff}}/{\delta x}=\beta{\partial   F}/{\partial x}=0$, where $x=V,t_{f},\lambda,\Delta$. 
In practice, we impose these conditions on our numerical calculations, thereby imposing four self-consistency conditions:
\begin{align}
&\frac{1}{2N{\cal N}_s}\sum_{\mathbf{k},\alpha ,\sigma}\gamma_\bk\langle f\dg_{\mathbf{k},\alpha,\sigma} f_{\mathbf{k},\alpha,\sigma}\rangle + \frac{4{t}_f}{J_H}=0, \label{JHrel}\\
&\frac{1}{2N{\cal N}_s}\sum_{\mathbf{k},\alpha,\sigma}2(c_x-c_y)\langle \tilde \sigma f_{-\mathbf{k},\alpha ,-\sigma} f_{\mathbf{k},\alpha,\sigma}\rangle + \frac{4{\Delta}}{J_H}=0, \end{align}
\begin{align}
&\frac{1}{2N{\cal N}_s}\sum_{\mathbf{k},\alpha,\sigma,\sigma'}\langle c\dg_{\mathbf{k},\alpha,\sigma}\phi(\bk)_{\alpha\sigma,\alpha\sigma'}f_{\mathbf{k},\alpha,\sigma'}
\rangle+\frac{{V}}{J_K}=0, \\
&\frac{1}{2N{\cal N}_s}\sum_{\bk,\alpha,\sigma}\langle f\dg_{\bk,\alpha,\sigma}f_{\bk,\alpha,\sigma} \rangle=\frac{1}{2}, 
\end{align}
where the normalizing prefactors derive from the $2{\cal N}_s$ $f$ sites and the $N$ spin components. 
The last condition is the constraint of a half-filled $f$ band, $\langle {\cal F}\dg \tau_3 {\cal F}\rangle = 0 $. 
%Once the hybridization field $ V$ turns on, the intersite f-electron hybridization becomes finite and by \eqref{JHrel}, this induces  finite dispersion of strength $t_f$.  
In general, once a finite hybridization $V$ develops in the normal state, this gives rise to a finite hybridization between $f$ electrons at different sites, which by \eqref{JHrel} induces a finite $t_f$. 
%{\color{red} What do we say about the f-hopping. Generically, once $V$ turns on, we have to impose $\delta F/\delta t_{ij}=0$, which will lead to a finite hopping of the f-states. I think we neglected this? }
%The expectation value represents the thermal average. The translational invariance ensures position and time independence.
%--------------------------------------------------
\section{Results}
\subsection{Normal state: Emergent Spin-Orbit Coupling}

We begin with a discussion of the normal state.  Here, the key point is that the absence of local inversion symmetry at the Cerium sites modifies the electrostatic environment of the atoms. This polarizes the atomic orbitals, giving them an ill defined parity which opens up previously forbidden mixing between even and odd parity states. We will see how this modifies the spin orbit coupling (SOC). To elucidate this effect, we examine the $f$ propagator obtained after integrating out the conduction electrons. This takes the form:
\begin{equation}\label{f_propagatorz}
    \mathcal{G}_f(\mathbf{k},i\omega_n)=[i\omega_n-h_\mathbf{k}\tau_3-\Sigma_f(\mathbf{k},i\omega_n)]^{-1}.
\end{equation}
%The last term in \eqref{f_propagatorz} is the self-energy of the $f$-electron which represents the effect of the hybridization with the conduction electrons. 
The $f$ electron self-energy $\Sigma_f(\bk,i\omega_n)$ is represented by the following diagram
\begin{equation} 
\Sigma_f(\mathbf{k},i\omega_n)
=
\raisebox{-0.56\height}{%
\includegraphics[width=0.28\columnwidth]{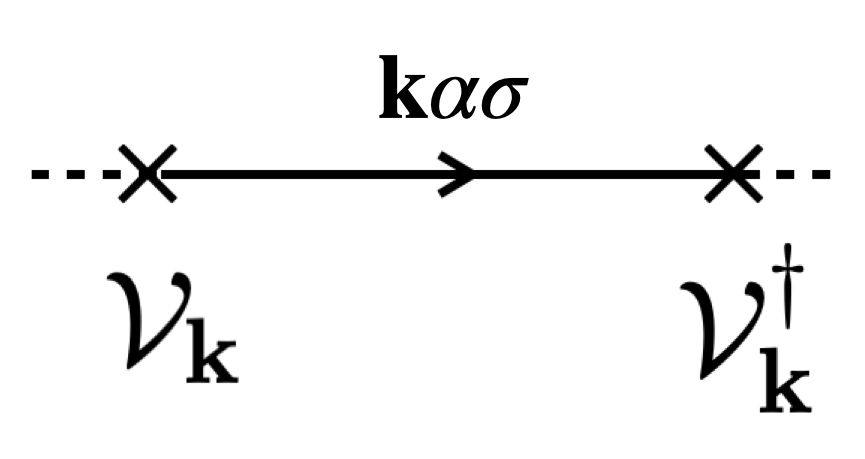}}
=
\mathcal{V}_{\mathbf{k}}
\frac{1}{i\omega_n-h_\mathbf{k}\tau_3}
\mathcal{V}_{\mathbf{k}}^\dagger,
\end{equation}
where ${\cal V}_{\bk} = V \phi(\bk)\tau_3$.
Let us now examine the spin orbit coupling generated by this process. The conduction electron dispersion $h_\mathbf{k}$ can be divided into sublattice independent and dependent parts,  $h_\mathbf{k}=h^0_{\bk}+{h}^1_{\bk}\alpha_1$. 
Now since $\tau_3$ and $\alpha_1$ are commuting operators, which can be simultaneously diagonalized, we can rewrite the 
conduction electron propagator in the form
\begin{equation}
    \frac{1}{i\omega_n-h_\mathbf{k}\tau_3}=\frac{(i\omega_n-h_\bk^0\tau_3)+{h}^1_{\bk}\alpha_1\tau_3}{(i\omega_n-h_\bk^0\tau_3)^2-(h^1_{\bk})^2}
\end{equation}
Reinstating $\alpha_1$ and $\tau_3$ as operators, the
self-energy takes the form 
\begin{widetext}
    \begin{align}\label{self_energy_expand}
    \Sigma_f(\mathbf{k},i\omega_n)
    =V^2 ({\bf s}_{\bk}\cdot \bm{\sigma}+iW\alpha_3\sigma_3)&\Bigg[ \frac{(i\omega_n-h_\bk^0\tau_3)+{h}^1_{\bk}\alpha_1\tau_3}{(i\omega_n-h_\bk^0\tau_3)^2-(h^1_{\bk})^2}
    \Bigg]({\bf s}_{\bk}\cdot \bm{\sigma}-iW\alpha_3\sigma_3).
\end{align}
\begin{align}
&=\frac{V^2 \tau_3}{(i\omega_n-h_\bk^0\tau_3)^2-(h_\bk^1)^2}
\biggl({(i\omega_n\tau_3-h_\bk^0)\biggl[({s}_{\bk}^2+W^2)-2 W\alpha_3 ({\bf \hat z} \times {\bf s}_{\bk})\cdot\bm{\sigma}\biggr]+h_\bk^1\alpha_1({{s}_{\bk}}^2-W^2)}\biggr).
\end{align}
\end{widetext}
where  ${\bf s}_{\bk}=(\sin k_x,\sin k_y)$ and ${s}_{\bk}^2 = \sin^2 k_x+\sin^2 k_y$.
% \begin{equation}\label{self_energy_expand}
% \begin{aligned}
% \Sigma_f(\mathbf{k},i\omega_n)=
% &\kappa V_0^2 (\sin(\mathbf{k})\cdot\vec{\sigma}+iW_3\alpha_3\sigma_3)(h_0+\vec{h}\cdot\vec{\alpha})\\
% &\times(\sin(\mathbf{k})\cdot\vec{\sigma}^*-iW_3\alpha_3\sigma_3).
% \end{aligned}
% \end{equation}
%where $\kappa =[(i\omega_n-h^0_{\bk})^2-\vec h_\mathbf{k}^2]^{-1}  $.
We see that the interference between the two terms in the hybridization generates staggered spin orbit coupling with a cross product structure between momentum and spin. At low energies, 
\begin{equation}\label{self_energy_SOC}
 \Sigma^{\rm SOC}_f(\mathbf{k},i\omega_n)=   \frac{2V\tilde W \alpha_3 \tau_3}{(i\omega_n-h_\bk^0\tau_3)^2-(h_\bk^1)^2} (s_y\sigma_1-s_x\sigma_2) ,
\end{equation}
where ${W}=\tilde {W}/V$ demonstrating that the self-energy of the $f$ electron in Eq.\ref{self_energy_SOC} develops a frequency dependent form of the ASOC interaction deduced from symmetry analysis. 
Unlike previous models which added the ASOC interaction as a phenomenological term in the normal state, here the  essential physics is produced from an interplay between the local parity breaking hybridization and the intersite ${\bf s}_{\bk}\cdot \bm{\sigma}=\sbk$ hybridization that carries the strong SOC effects of the $f$-states. The new local parity breaking term $i\alpha_3W\sigma_3$ in the hybridization form factor $\phi_{\mathbf{k}}=({\bf s}_{\bk}\cdot \bm{\sigma}+i{W}\alpha_3\sigma_3)$ provides the spin flip necessary to swap the spin components of the intersite hybridization $\sin \mathbf k_\perp\bm \cdot\bm\sigma$ generating an antisymmetric SOC. 
The frequency dependent prefactor encodes the virtual processes of the $f$ electrons fluctuating into  the conduction sea and returning  having acquired the necessary spin and momentum symmetry required to induce the ASOC interaction. In otherwords, Kondo screening in a noncentrosymmetric environment creates a heavy Fermi liquid  with the necessary ingredients for a triplet superconducting state.

The normal state properties can be seen in how the dispersion and Fermi surfaces change across the Brillouin zone.
\begin{figure}
    \centering
    \includegraphics[width=1.0\linewidth]{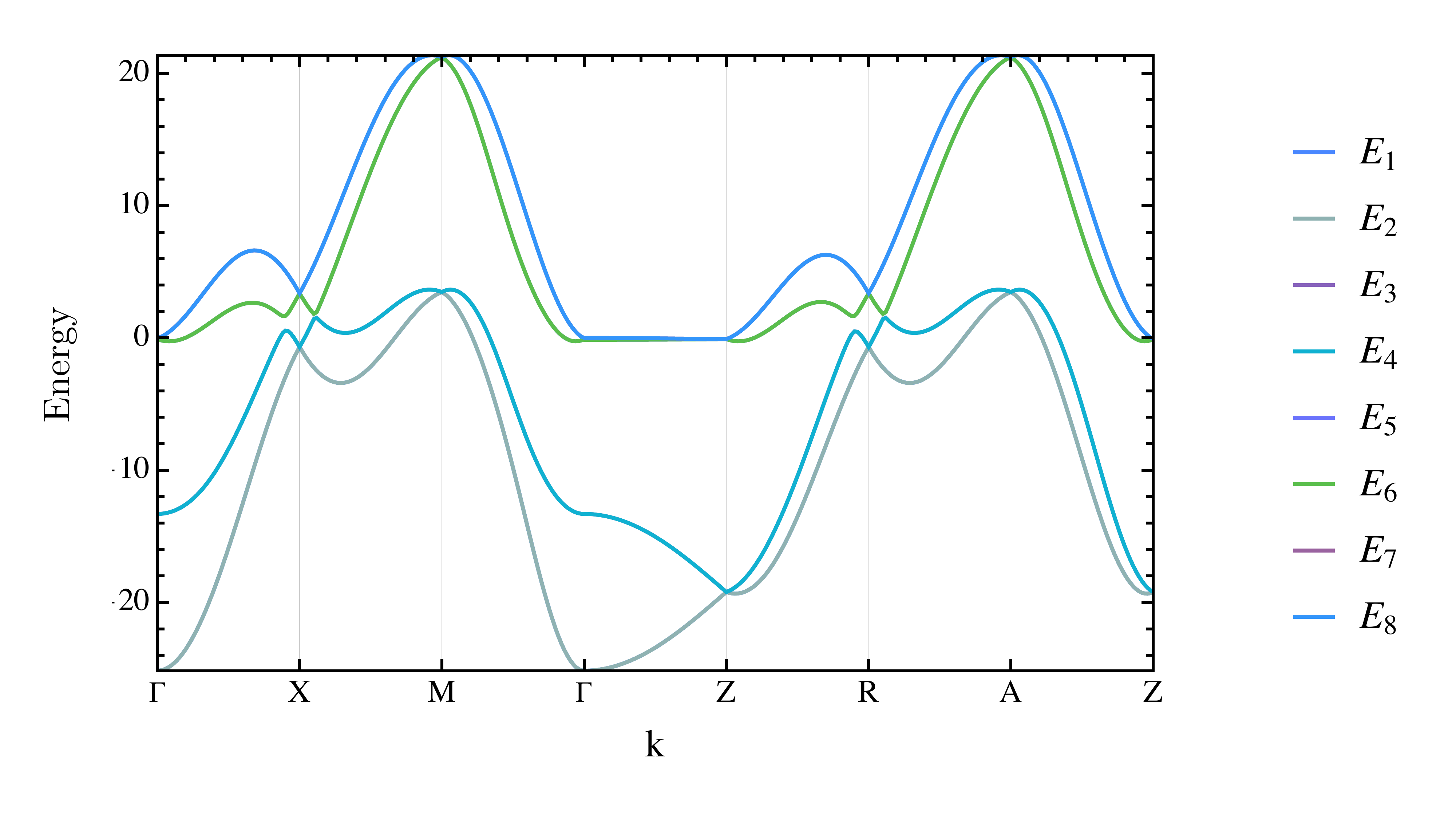}
    \caption{The normal state bandstructure plotted along the high symmetry points. The intralayer hopping is given by $t=5$, the interlayer hopping $t_s=1.5$ and the f electron hopping is $t_f=0.5$, the chemical potential $\mu=-1$ and the f electron constraint $\lambda=-1.7$, the hybridization strengths are $V=3$ and $W=2$. The doubly degenerate bands are ensured by Kramer's degeneracy.}
    \label{fig:dispersion_normal}
\end{figure}
The dispersion plot of the normal state is shown in Figure \ref{fig:dispersion_normal}. The dispersion plotted along the path given by the high symmetry points  $\rm \Gamma=(0,0,0),\,X=(\pi,0,0),\, M=(\pi,\pi,0), Z=(0,0,\pi),R=(\pi,0,\pi),A=(\pi,\pi,\pi)$.
\begin{figure}
    \centering\hspace*{-0.0\columnwidth}
    \includegraphics[width=1.\linewidth]{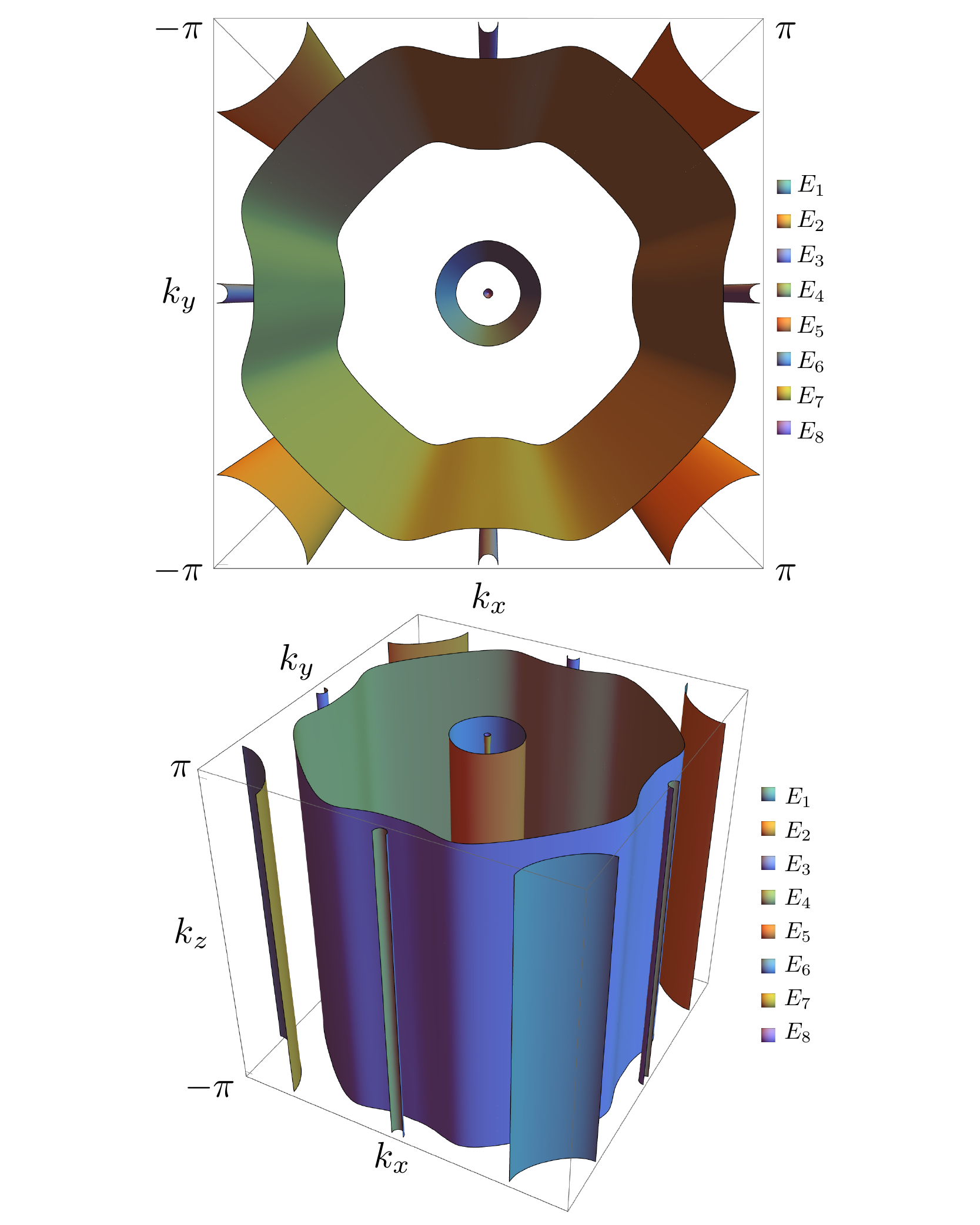}
    \caption{The 3D Fermi surface of the normal state, the surfaces are doubly degenerate. The parameters for the normal state are given by the intralayer hopping $t=5$, the chemical potential $\mu=-1$ the interlayer hopping $t_s=1.5$ and the $f$ electron hopping $t_f=0.5$, and the $f$ electron chemical potential $\lambda=-1.7$, the hybridization strengths are $V=3$ and $W=2$. Top: the top view. Bottom: the side view. }
    \label{fig:FS_3D}
\end{figure}
The 3D fermi surfaces of the normal state are shown in Figure \ref{fig:FS_3D}.

\subsection{Superconducting State}
\begin{figure}
    \centering
    \hspace*{-0.1\columnwidth}%
    \includegraphics[width=1.1\columnwidth]{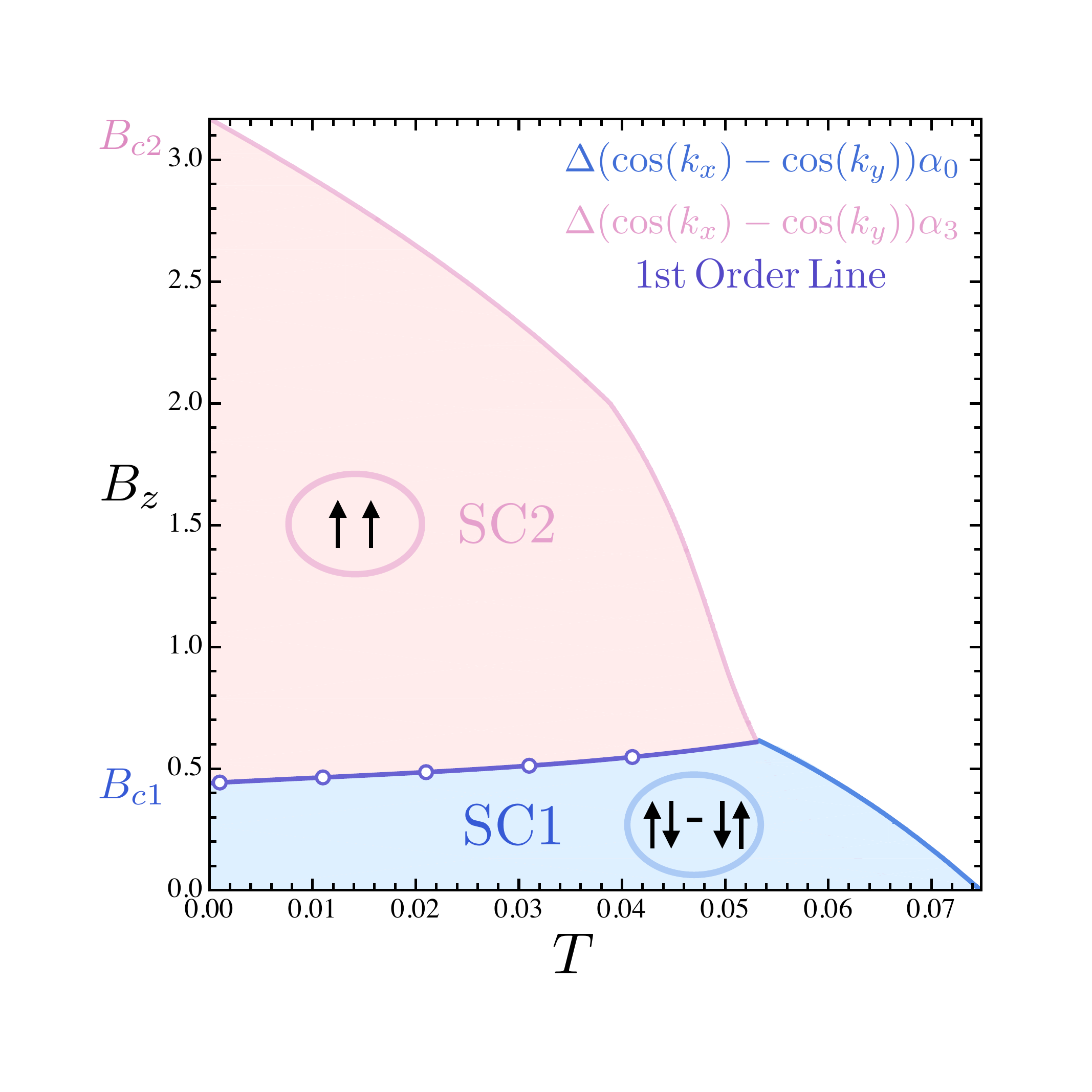}
    \caption{The superconducting phase diagram of applied field $B_z$ along the $z$-axis w.r.t $T$ for uniform $\Delta_\mathbf{k}\alpha_0$ and staggered $\Delta_\mathbf{k}\alpha_3$ phases. The SC1 phase is given by the uniform state which is Pauli limited and the high field SC2 phase is given by the staggered state. The $T_c$ was calculated by computing the pair susceptibility $\chi_P(B_z,T_c)$ obtained from taking the second derivative w.r.t pairing field $\Delta$. The first order phase transition line is shown in purple which separates the two second order phase transitions.}
    \label{fig:field_Tc}
\end{figure}

We now examine the superconductor, and the effect of a magnetic field in stabilizing
an odd-parity condensate. 
In the absence of a magnetic field, the even parity solution with its lower kinetic energy, is expected to be more stable, exhibiting a higher $T_c$ than its odd parity counterpart. Interestingly, in the limit where the hybridization is a constant, $\mathcal{V}_\mathbf{k}=V_0$, there is no staggered solution. The only solution is the uniform pairing which leads to the conventional $d$-wave singlet pairing within the layers commonly seen in the 115 heavy fermion superconductors \cite{heavy_fermion_SC_115},\cite{115}. In \ce{CeRh2As2}, neutron scattering does indeed detect a peak at the dynamic structure factor at $\mathbf{Q}=(\pi,\pi)$, a hallmark for AFM fluctuations \cite{Broholm24}.

To investigate the effect of the applied field on the superconducting state, we now switch on a Zeeman field as defined in \eqref{scham}. Although \ce{CeRh2As2} is a type II superconductor which develops a vortex lattice in a field,  the effect of spatial modulation is assumed to be negligible for the low temperature and high field regime where Pauli limiting dominates, as described in \eqref{scham}.

Application of a Zeeman field affects the critical temperature $T_c$ which can be computed from the pair susceptibility $\chi_P$. Generally, in the normal state, the pair susceptibility enters into the Landau free energy functional via the quadratic term $F[\Delta]=|\Delta|^2(\chi_P(B_z)-2/J_H)+\beta|\Delta|^4...$ calculated by taking the second derivative w.r.t the pairing field $\Delta$. At $T_c$, the $\partial ^2 F/\partial \Delta ^2|_{\Delta =0}=0$ giving the well known criteria for $T_c$:
\begin{equation}\label{pair_susceptibility}
\chi_P(B_z,T_c)-\frac{2}{J_H}=0.
\end{equation}
In our calculations, the pair susceptibility was obtained by evaluating the second derivative of the Free energy, $\chi_P(B_z,T_c)-2/J_H=-{\partial^2 F(B_z,T_c)}/{\partial\Delta^2}$ for a range of magnetic field strengths for both uniform and staggered solutions. The corresponding phase boundaries for the uniform $\Delta_{\mathbf{k}}^{(0)}$ and staggered $\Delta_{\mathbf{k}}^{(3)}$ solutions are shown in Figure.\ \ref{fig:field_Tc}. The first order SC1 to SC2 boundary was determined by comparing the fully evaluated mean-field free energies of the self-consistent uniform and staggered superconducting solutions at the same $(B_z,T)$. The transition occurs where these two free energy branches cross; the pair susceptibility criterion in Eq.~\eqref{pair_susceptibility} determines the normal state instability lines, not the first order boundary. The uniform state (SC1) has a higher condensation energy than the staggered state since staggering the phase of the pairing field $\Delta$ requires a higher kinetic energy \cite{parity_mixed_noncentrosymmetric}, and this results in a higher $T_c$ at zero field. However, the condensation energy of the singlet SC1 phase is more strongly susceptible to a Zeeman splitting, ultimately raising its energy above the staggered, triplet SC2 phase. Once a magnetic field is applied, there is a first order phase transition to the staggered state. There is a thus field induced parity switching between the uniform even parity state to the staggered odd parity state providing a concrete Kondo lattice demonstration of the field induced parity switching behavior observed in experiments.
\begin{figure}
    \centering\hspace*{-0.15\columnwidth}
    \includegraphics[width=1.3\linewidth]{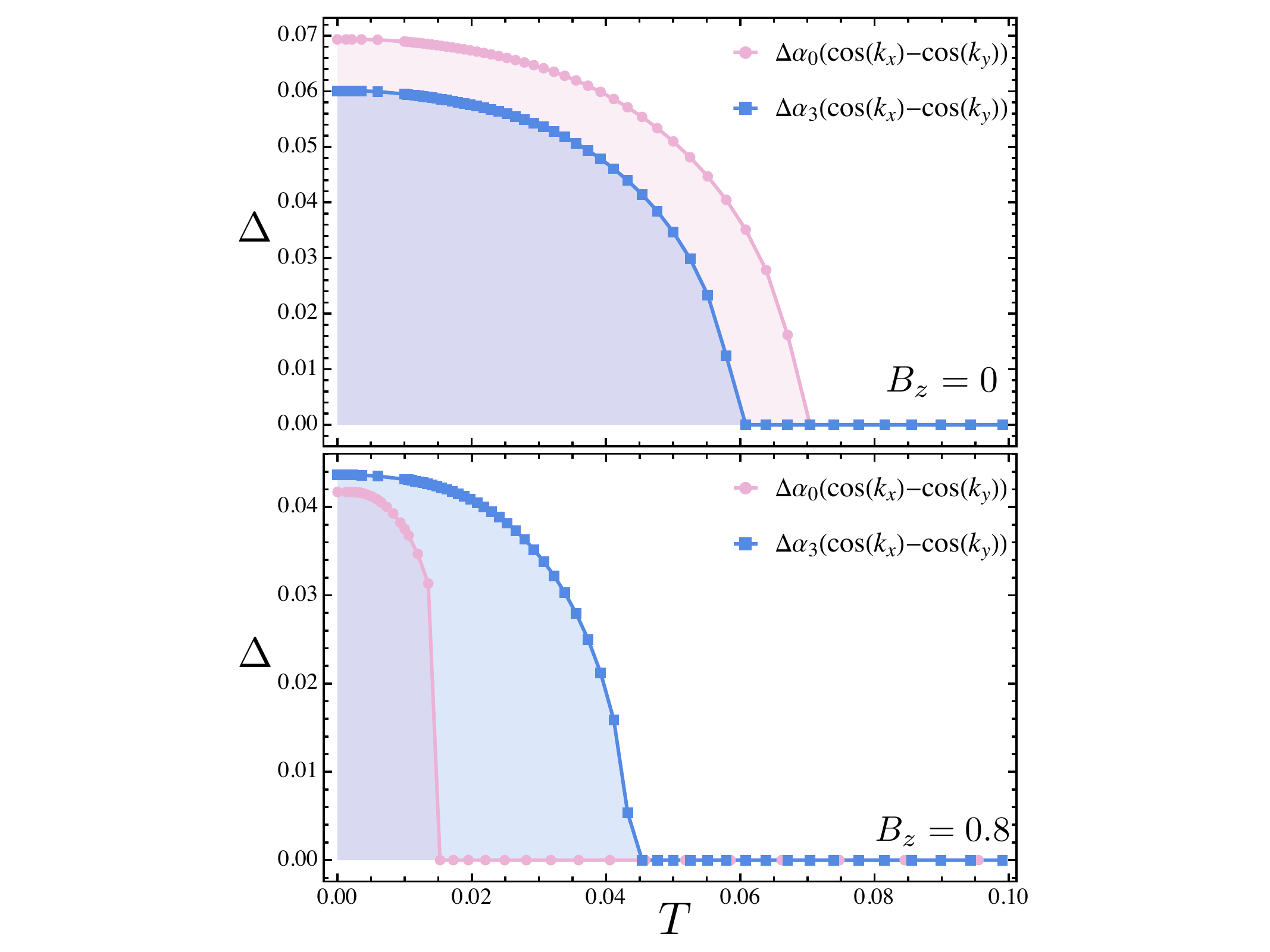}
    \caption{The phase diagram of the pairing gap vs temperature for both uniform $\Delta_{\mathbf{k}}\alpha_0$ and staggered $\Delta_{\mathbf{k}}\alpha_3$ phases for different magnetic field strengths $B_z$. The calculation of the gap magnitude $\Delta$ was determined self consistently. Top: $B_z=0$ dominated by uniform pairing $\Delta_{\mathbf{k}}\alpha_0$ state. Bottom: $B_z=0.8$ is dominated by staggered $\Delta_{\mathbf{k}}\alpha_3$ state.}
    \label{fig:gap_T}
\end{figure}

For further investigation, a self consistent calculation was performed to analyze the uniform and staggered pairing states. The heavy Fermi liquid phase has already been developed and frozen with the Kondo hybridization parameters $V=3$ and $W=2$ which are the intersite and onsite hybridization respectively and the Kondo and Heisenberg coupling strengths are set to $J_K=0.3$ and $J_H=0.15$ respectively. The intralayer and interlayer hopping strengths are given by $t=5$ and $t_s=1.5$ respectively and the chemical potential was set to $\mu=-1$. The hopping scales are consistent with a quasi-2D structure. The $f$ electron hopping strength is set to $t_f=0.5$ with $t_f\ll t$ and the chemical potential for the $f$ electron was set to $\lambda=-1.7$ with a $q=1/3$ charge. The qualitative results do not depend on the microscopic details of the band structure as long as keeping the ratio of $t/t_s> 1$ for the interlayer to intralayer coupling and $VW/t_s>\ 1$ for the interlayer to hybridization strength. With these parameters, a Newton Raphson method was used to find the saddle point of the pairing field by minimizing the Free energy w.r.t the pairing field $\Delta$ for the uniform and staggered solutions. The resulting phase diagram of the pairing gap $\Delta$ vs temperature in the presence of magnetic field is given by Fig.\ref{fig:gap_T}. 

The zero field behaviour of the uniform and staggered states is shown in the top plot of Fig.\ref{fig:gap_T} which shows the pairing field and $T_c$ for the two states are nearly degenerate. The uniform state has a higher $T_c$ emphasising the dominance of the even parity singlet state at zero field. However, once a magnetic field of $B_z=0.8$ is applied, as shown in the lower plot Fig.\ref{fig:gap_T}, the even parity state is suppressed by Pauli limiting effect. When the $T_c$ of the staggered odd parity state starts to overcome the uniform even parity state, a field induced transition takes place. The staggered odd parity state is more weakly susceptible to Pauli limiting effects.
%, but is still limited from to orbital depairing effects. 
The upper critical field of the staggered state is $B_{c2}\approx 3.2$ for which the uniform state completely vanishes. To understand how a spin singlet pairing in the layers staggered could realise a triplet odd parity state, the pairing gap will be projected onto the low energy pseudospin basis.  

\subsection{Projection of the Gap onto pseudospin basis}
\begin{figure}[ht]
    \centering\hspace*{-0.\columnwidth}
    \includegraphics[width=1.\columnwidth]{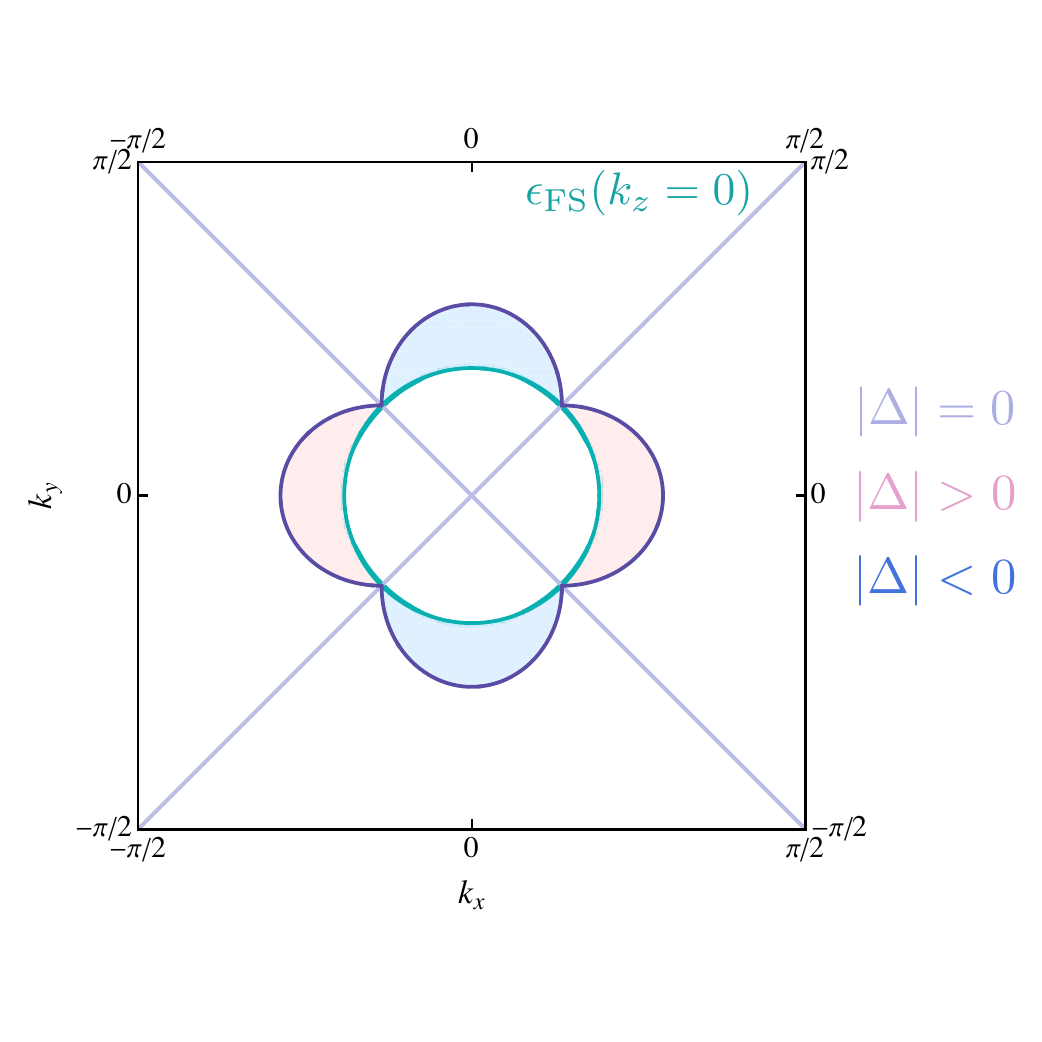}
    \caption{Uniform gap $\Delta_{\bk}^{(0)}$ projected onto the normal state basis which reflects a diagonal projection where the gap is given by $\psi_0({\mathbf{k}})$. The Fermi surface contour is shown by the green. The diagonal lines in light purple are the zeros of the projected gap function $\Delta_{\bk}^{(0)}=0$ which represent a B$_{1g}$,  $d_{x^2-y^2}$ nodal structure. The pink and blue shaded regions show the sign of the pairing gap corresponding to positive $\Delta_{\bk}^{(0)}>0$ and negative $\Delta_{\bk}^{(0)}<0$ gaps respectively. The purple contour is the angular dependence of the projected gap showing vanishing gap near the nodes.}
    \label{fig:even_parity}
\end{figure}
\begin{figure}[ht]
    \centering\hspace*{-0.1\columnwidth}
    \includegraphics[width=1.2\columnwidth]{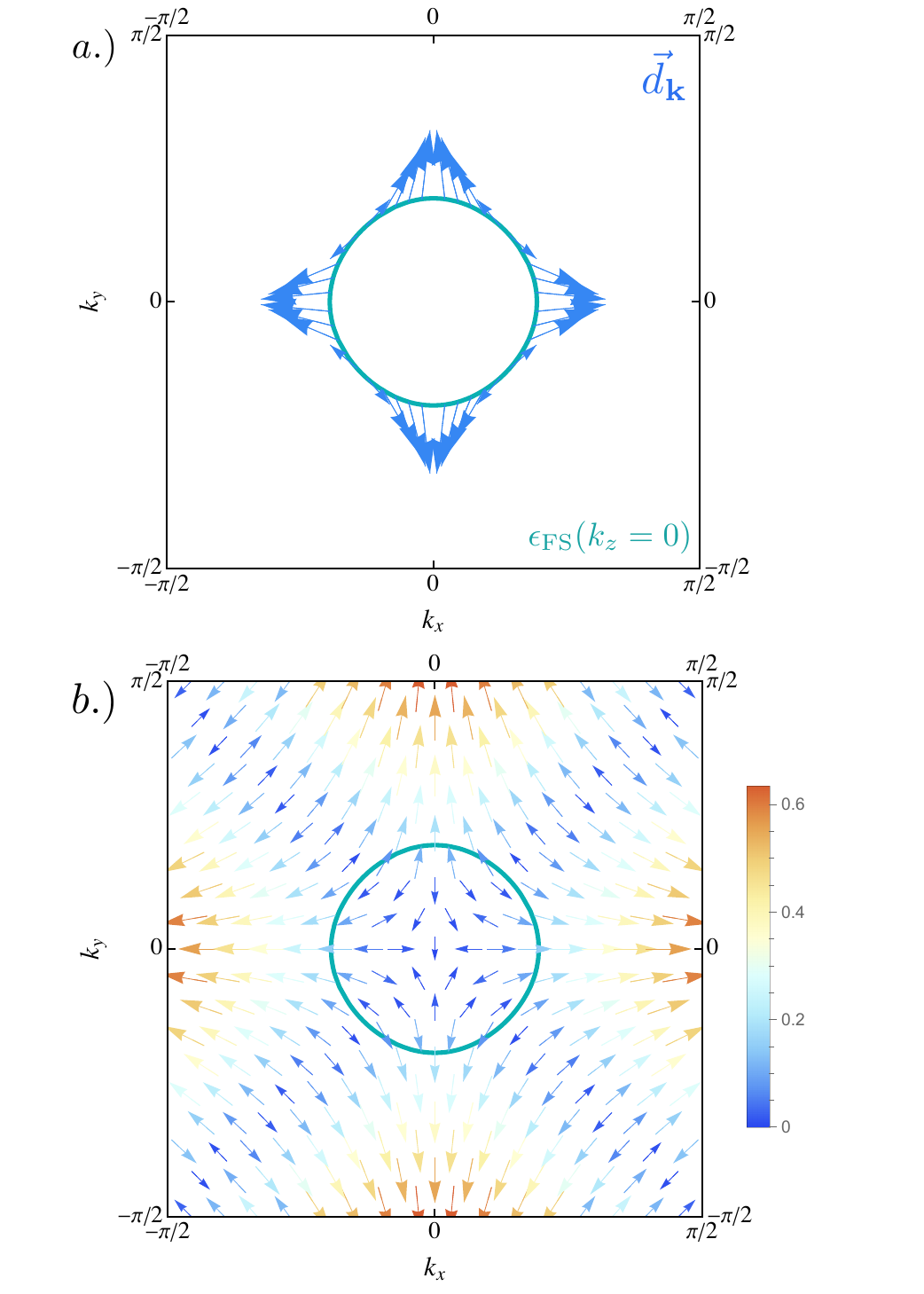}
    \caption{Plots displaying the B$_{1u}$ triplet  gap function, showing its product structure, $\Delta_{\downarrow \downarrow}(\bk) = d_x(\bk)+ i d_y(\bk) \propto \cos (2\theta) e^{-i\theta}$  which vanishes along the diagonals and is a mixture of $f$ and $p$ wave pairing. {\bf a.} The plot of the projected staggered gap $\Delta_{\bk}^{(t)}$ onto the normal state basis. To more clearly display the d-wave form factor, the Rashba d-vector has been rotated anticlockwise by 90$^\circ$, corresponding to a gauge transformation $\Delta\rightarrow -i \Delta$.   The Fermi surface is shown by a green contour. {\bf b.} the   $\vec d_\bk$ vector across the Brillouin zone.   }
    \label{fig:odd_parity_triplet}
\end{figure}

\begin{comment}
\begin{figure}[ht]
    \centering\hspace*{-0.1\columnwidth}
    \includegraphics[width=1.2\columnwidth]{Figures/odd_parity_triplet_pseudo.pdf}
    \caption{Plots displaying the triplet gap function, showing its product structure, $\Delta_{\downarrow \downarrow}(\bk) = d_x(\bk)+ i d_y(\bk) \propto \cos (2\theta) e^{-i\theta}$  which vanishes along the diagonals and is a mixture of $f$ and $p$ wave pairing. {\bf a.} The plot of the projected staggered gap $\Delta_{\bk}^{(3)}$ onto the normal state basis. In this plot, the spin basis $\sigma_i$ was also projected onto the Fermi surface as $ \xi^{\sigma\dagger}_{\bk n} \sigma_i\xi^{\sigma'}_{{\bk}n}\rightarrow \sigma'_i$. The Fermi surface shown by a green contour. {\bf b.} the   $\vec d_\bk$ vector across the Brillouin zone. }
    \label{fig:odd_parity_triplet_pseudo}
\end{figure}
\end{comment}
To understand the nature of the gap structure, the gap must be projected onto the Fermi surface.  The procedure requires the gap operator originally in the high energy sublattice and spin subspace to be evaluated within the Kramers pair of energy eigenstates in a given band \cite{nonsymmorphic}
\begin{equation}\label{extract}
\Delta(\bk)_{\sigma\sigma'} = 2 \Delta(c_x-c_y)\ \bigl[\xi^{\sigma\dagger}_{\bk n} \cdot \left(\tau_1\alpha_m P_f\right) \cdot\xi^{\sigma'}_{{\bk}n}\bigr].
    %U_{\mathbf{k}\sigma n}^\dagger(\Delta_{\mathbf{k}\sigma\sigma'}\kappa_{f}\otimes\alpha_m)U_{\mathbf{k}\sigma'n'}\rightarrow\Delta_{\mathbf{k},nn'}^{(m)}
\end{equation}
Here $\xi^{(\sigma)}_{\bk n}$, ($\sigma \in ( \uparrow, \downarrow)$ 
denotes the Kramers doublet of eigenstates of the superconducting mean field Hamiltonian at momentum $\bk$ in the $n$-th band and $P_f$ projects into the $f$ sector. The sublattice matrices can take either $\alpha_m =\alpha_0$ or $\alpha_3$ which denotes the sublattice for the uniform, and staggered gap functions respectively. 

To consistently calculate this quantity, we need to consistently fix the phase relationship between the pseudospin up and down eigenstates. We calculate the up eigenstate by a applying a small field to split the Kramers degeneracy, selecting it as the ``up" state as the lowest energy member of the doublet.  We then apply  time-reversal and parity inversion to the up state to generate its down-spin partner,  $\xi^{\downarrow} _{\bk n} = i \alpha_1\sigma_2  (\xi^{\uparrow} _{\bk n})^*$.

The Fermi surface projection of the uniform gap $\Delta_{\mathbf{k}}\alpha_0$ is shown in Fig.\ref{fig:even_parity}. As expected from \eqref{extract}, the gap has zeros along the $k_x=k_y$ derived from the in-plane $d$-wave gap structure. The colors show the sign of the gap  $\text{sgn(}\Delta)$ where the $+$ and $-$ signs are denoted by pink and blue respectively. Thus the projected gap structure of the uniform gap on the Fermi surface is an even parity $d$-wave spin singlet $\Delta_{\mathbf{k}n}^{(0)}$ where  $n$ represents the band index
\begin{equation}
    \Delta_{\mathbf{k}}^{(s)}=\begin{pmatrix}
        \psi_0(\mathbf{k}) & 0\\
        0 & \psi_0(\mathbf{k})\\
    \end{pmatrix}.
\end{equation}
Here $\psi_0(\mathbf{k})$ is an even, generally signed gap function. Throughout this subsection, $\Delta_{\bk}$ denotes the reduced gap matrix, while the full anomalous gap matrix is $\hat\Delta_{\bk}=\Delta_{\bk}(i\sigma_2)$ in the pseudospin basis.

\subsection{The Symmetry of the singlet and pseudo-spin triplet states}

\subsubsection{Symmetry of the gap}
We now discuss the symmetry and physical origin
of the spin-orbit generated pseudospin triplet state. 
The projection of the staggered gap $\Delta_{\mathbf{k}}^{(t)}$ is shown in Fig.\ref{fig:odd_parity_triplet}. 
%There appears to be no zeros in the projected staggered gap but shows a fully gapped structure with a spin texture with a $\vec{d}_\mathbf{k}$-vector. 
The key result is that even though the in plane pairing is singlet, the staggered pair potential on alternating layers transforms it into an odd parity triplet state on the Fermi surface. The projected odd-parity pseudospin-triplet gap, with the universal factor $i\sigma_2$ suppressed, has the structure 
\begin{align}
    \Delta_{\mathbf{k}}^{(t)}={\vec{d}_\mathbf{k}\cdot\vec{\sigma}} = 
    \begin{pmatrix}
          0 & d_{-}(\bk) \\
          d_{+}(\bk) & 0\\
    \end{pmatrix}
\end{align}
where $d_{\pm}(\bk) = d_x(\bk)\pm i d_y(\bk)$, 
%Where the d-vector $\vec{d}_\mathbf{k}=(d_x(\mathbf{k}),d_y(\mathbf{k}),0)$ vector represents the spin components. 
If we rewrite the pairing Hamiltonian in terms of the quasiparticle creation operators,  $a^\dagger_{\bk,\sigma}$, 
\begin{eqnarray} 
a^\dagger_{\bk, \sigma} [\vec{d}_\mathbf{k}\cdot\vec{\sigma} (i\sigma_2)]_{\sigma, \sigma'}a^\dagger_{-\bk,\sigma'}&=& \cr
(d_{+}(\bk) a^\dagger_{\bk ,\downarrow}a^\dagger_{-\bk, \downarrow} -&& \hskip - 5mm d_{-}(\bk) a^\dagger_{\bk ,\uparrow}a^\dagger_{-\bk ,\uparrow}),
\end{eqnarray}
Thus the off-diagonal structure of the reduced gap matrix corresponds, after multiplication by $i\sigma_2$, to diagonal equal pseudospin pairing amplitudes $\hat\Delta_{\uparrow\uparrow}=-d_-$ and $\hat\Delta_{\downarrow\downarrow}=d_+$. These are pseudospin labels, not necessarily the physical electron spin quantum numbers. Fig. \ref{fig:odd_parity_triplet} shows the variation of $\vec{d}_\mathbf{k}$ around the Fermi surface.
To a good approximation $\Delta_{\downarrow \downarrow}(\bk) = d_x(\bk)+ i d_y(\bk) \propto i \cos (2\theta) e^{-i\theta}$
is a product of a $d$-wave and $p$-wave gap function. We can decompose this into  $ \Delta_{\downarrow \downarrow}(\bk)\propto   (e^ {i\theta} + e^{-3i\theta})$, corresponding to angular harmonics of winding numbers $+1$ and $-3$ (often described as $p$- and $f$-wave components); the crystalline representation is determined by the full tetragonal symmetry.  
In the tetragonal point group approximation, the uniform intralayer gap $\cos k_x-\cos k_y$ belongs to $B_{1g}$ of $D_{4h}$. For the projected staggered state, the approximate form $\vec d_{\bk}\propto(\cos k_x-\cos k_y)(\sin k_y,-\sin k_x,0)$ is odd under inversion, changes sign under a fourfold rotation, and is even under a vertical mirror when the axial vector transformation of $\vec d$ is included. It therefore transforms as $B_{1u}$ in this pseudospin convention. This identification is for the displayed leading harmonic; a complete space group classification must also account for the nonsymmorphic sublattice representation and the momentum dependent pseudospin gauge. The interesting feature of the $\vec{d}_\mathbf{k}$ vector is that it lies in the basal plane, and this makes the resulting triplet state robust against fields along the $z$-axis.
This projected symmetry is consistent with the analysis of Ishizuka
\textit{et al.}~\cite{Ishizuka2024}, who classified candidate
superconducting gaps in \ce{CeRh2As2} using electronic structure
calculations and crystalline symmetry. For a magnetic field along the
$c$ axis, the relevant point group classification is reduced to
$C_{4h}$; the leading $B_{1u}$ gap identified above belongs to its
$B_u$ sector. In particular, their staggered spin singlet $B_u$
basis includes $k_x^2-k_y^2$, the long wavelength limit of our
$\cos k_x-\cos k_y$ form factor. Our projected gap
$\vec d_{\bk}\propto(\cos k_x-\cos k_y)
(\sin k_y,-\sin k_x,0)$ provides a higher harmonic realization
of the same $B_u$ symmetry class. The diagonal zeros of this leading
form factor need not be enforced for every gap in that representation.
The approaches are complementary: their work establishes crystalline
symmetry constraints and possible topological properties, whereas our
Kondo lattice calculation provides a microscopic mechanism for the
emergence of the projected pseudospin triplet structure. This
agreement at the point group level does not replace a full
nonsymmorphic analysis of our heavy band gap.

\subsubsection{Physical origin of the triplet gap}

The origin of this pseudospin triplet structure
can be understood physically by considering the bonding and
antibonding combinations of $f$ states on neighboring
Ce layers.
In the absence of staggered spin orbit coupling,
the heavy quasiparticle bands are approximately
eigenstates of the interlayer hopping matrix
$\alpha_1$, with eigenvalues $\pm1$.
Since $\{\alpha_1,\alpha_3\}=0$, the staggered
pairing and staggered spin-orbit coupling,
both proportional to $\alpha_3$, are off-diagonal
in this approximate band basis.

Virtual transitions between the low energy Fermi
surface and higher energy quasiparticle bands
then generate an effective intraband pairing
interaction.
To leading order, this contribution has the
symmetry of the product of the staggered
$d$-wave pairing and the staggered Rashba
spin orbit coupling:
\begin{equation}
\Delta_{\rm eff}(\bk)
\sim
\frac{WV\Delta}{\delta E_{\bk}}
(c_x-c_y)
(s_y\sigma_x-s_x\sigma_y),
\end{equation}
where $\delta E_{\bk}$ denotes the characteristic
energy separation between the bands.
The two staggered factors combine through
$\alpha_3^2=1$, producing a sublattice even,
momentum odd pseudospin triplet gap.

This mechanism illustrates how an odd-parity
pseudospin triplet superconducting state can emerge
from staggered spin singlet pairing in the presence
of locally broken inversion symmetry.
More generally, analogous processes may occur in
multi sublattice superconductors where staggered
pairing interacts with staggered spin orbit coupling.

\section{Conclusion}
In this work we have developed a Kondo lattice framework for the field induced parity transition for the heavy fermion superconductor \ce{CeRh2As2}. Our central result combines the Kondo hybridization and local inversion symmetry breaking which provides a microscopic mechanism by which the two sublattice structure of \ce{CeRh2As2} can generate nearly competing superconducting states of opposite parity. Local inversion symmetry breaking at the Ce sites generates a new onsite spin flip hybridization which combines with the non local odd parity hybridization to generate a staggered antisymmetric spin orbit coupling (ASOC). For pairing driven by antiferromagnetic exchange interaction within the Ce layers, the uniform and staggered solutions are then determined by how they transform under inversion, which are even and odd respectively. In this model, the uniform intralayer $d$-wave pairing produces the low field, even parity state, while the staggered intralayer $d$-wave pairing with opposite signs of the pairing gap between the two Ce sublattices produces the high field, odd-parity pseudospin-triplet state after projection onto the heavy quasiparticle bands. A magnetic field along the $c$ axis suppresses the uniform phase much more strongly than the staggered phase leading to the field induced phase transition between the two superconducting phases.

Projection onto the normal state describing the heavy Fermi surface further shows an important consequence of the two pairing states. Although the underlying interaction was decoupled through an intralayer spin singlet $d$-wave pairing, the staggered state projected onto the Fermi surface reveals odd parity pseudospin triplet gap structure with an in-plane $\vec d_{\mathbf{k}}$ vector. The resulting odd parity state is not affected by the $c$-axis magnetic field as the $\vec d_\bk\perp B_z$. More generally, this illustrates that in a locally noncentrosymmetric Kondo lattice system, the spin, sublattice and parity structure of the microscopic pairing mechanism cannot be thought of independently. The interplay between the Kondo hybridization and ASOC can modify the structure of superconducting gap function when it is projected onto the normal state heavy quasiparticle basis. The key physics here is not to identify the particular $d$- or $p$-wave pairing interaction within the layers, but rather a mechanism by which the locally broken inversion symmetry with the two sublattice structure leads to a closely competing superconducting states with opposite global parity.

This distinction is particularly important because the microscopic pairing symmetry of \ce{CeRh2As2} has not yet been established experimentally. We have assumed that antiferromagnetic exchange favours an in-plane $d$-wave singlet pairing, but an intrinsic spin triplet pairing interaction is also possible. Indeed, a Hund's--Kondo mechanism has been proposed in which valence fluctuations into high spin configurations generate triplet correlations, with the two sublattice structural motif allowing these correlations to acquire an odd parity form factor on the heavy Fermi surface \cite{Hazra_Coleman_2023}. Starting from an intralayer triplet pairing rather than a singlet state would alter the correspondence between the uniform/staggered sublattice structure and the parity of the superconducting state, while retaining the central importance of the locally noncentrosymmetric environment. It is therefore an important open problem to repeat the present microscopic analysis allowing singlet and triplet pairing channels to compete on an equal footing. Measurements capable of determining the spin susceptibility and gap anisotropy in the two superconducting phases would be especially valuable in distinguishing the different possibilities.

A second direction is suggested by the recently discovered van der Waals heavy fermion material \ce{CeSiI}. \ce{CeSiI} provides a striking two dimensional realization of a Kondo lattice. Within a single van der Waals layer, two triangular Ce layers surround a conducting silicene layer, and heavy fermion coherence and magnetic order persist into the ultrathin limit \cite{Posey2024CeSiI,Vijayvargia2024CeSiI}. More recently, pressure has been shown to suppress the antiferromagnetic state and produce a superconducting dome, establishing \ce{CeSiI} as a two dimensional heavy fermion superconductor \cite{Shi2026CeSiI}. Its two Cerium layer structure that naturally leads to the question whether a sublattice dependent Kondo hybridization can generate an analogous relation between local symmetry, ASOC and superconducting parity. Unlike \ce{CeRh2As2}, however, the Ce ions form triangular layers surrounding a conducting silicene layer, so the microscopic hybridization form factors and crystal symmetry are distinct. Extending the current Kondo lattice framework to the actual symmetry and electronic structure of \ce{CeSiI} would therefore test which aspects of the parity switching mechanism are generic to a two sublattice Kondo lattice system and which are specific to the noncentrosymmetric \ce{CeRh2As2}. The ability to exfoliate, gate and assemble \ce{CeSiI} into van der Waals heterostructure additionally raises the prospect of controlling the relative Kondo coupling of the two Ce layers and hence tuning the balance between competing superconducting states.

An equally revealing comparison is provided by $\beta$-\ce{CeNi2As2}. Recent work has established superconductivity in the CaBe$_2$Ge$_2$ isomorph of \ce{CeNi2As2}, which is a structural analogue of \ce{CeRh2As2} and possesses the same essential locally noncentrosymmetric two sublattice structural motif \cite{Morineau2026CeNi2As2}. Yet $\beta$-\ce{CeNi2As2} does not display the characteristic field induced high field phase observed in \ce{CeRh2As2}. This comparison to \ce{CeRh2As2} is important. It shows that the locally noncentrosymmetric crystal structure by itself is not sufficient to a field induced parity transition. Within the framework developed here, the occurrence of such a transition also depends on quantitative microscopic ingredients. These are the strength and momentum dependence of the Kondo hybridization, the resulting staggered ASOC, the interlayer coupling, the structure of the heavy Fermi surface, the pairing interaction, and the degree of nearly degenerate uniform and staggered superconducting states. A microscopic comparison of \ce{CeRh2As2} and $\beta$-\ce{CeNi2As2} could provide a stringent test of the mechanism proposed here. Rather than asking only why \ce{CeRh2As2} has the high field phase, one can ask which microscopic ingredient present in \ce{CeRh2As2}, but absent or weaker in $\beta$-\ce{CeNi2As2}, is responsible for the high field phase.

A further challenge is to establish the superconducting symmetry classification at the full nonsymmorphic space group level. The assignments discussed here are based on the tetragonal point-group transformation properties of the leading gap form factors and their projected pseudospin structure. A complete analysis for $P4/nmm$ must additionally track the fractional translations, the sublattice dependent Bloch phases, and the momentum dependent symmetry sewing matrices of the heavy quasiparticle bands, especially near Brillouin zone boundaries. Such a treatment would test the robustness of the proposed irreducible representation assignments and clarify any symmetry enforced constraints on the gap nodes.

There remain further unresolved questions within \ce{CeRh2As2} itself. The superconductivity develops in close proximity to an additional ordered phase, Phase I whose microscopic origin remains uncertain, and antiferromagnetic order has also been reported at temperatures below $T_c$ \cite{cerh2as2_pressure,cerh2as2_AFM_order,cerh2as2_coexistence_field,cerh2as2_coexistence_muSR}. Understanding how these orders couple to the two superconducting phases may be important for a quantitative theory of the phase diagram. Similarly, the anomalous normal state transport and indications of a non Fermi liquid behavior suggests that the superconductivity develops from a strongly correlated normal state close to quantum criticality \cite{non_FL_cerh2as2}. These effects lie beyond the static mean field treatment used here and motivate extensions in which the Kondo coherence, magnetic fluctuations and superconducting instability are treated on an equal footing.

Our results suggest a broader perspective for locally noncentrosymmetric heavy fermion superconductors. The essential ingredient identified here is the interplay between Kondo hybridization with the crystal structure where the inversion exchanges two locally noncentrosymmetric magnetic atoms. This structure allows local parity breaking to become encapsulated as a staggered ASOC of the heavy quasiparticles and provides a natural route to competing superconducting states of opposite global parity. \ce{CeSiI} and $\beta$-\ce{CeNi2As2} now provide complementary tests of this idea.  \ce{CeSiI} extends the problem toward a tunable two dimensional Kondo lattice, while the $\beta$-\ce{CeNi2As2} enables an isostructural superconducting control system in which the high field phase is absent. Together with a systematic treatment of intrinsic triplet as well as singlet pairing, these materials should help determine whether the mechanism presented here is peculiar to \ce{CeRh2As2} or forms part of a more general organizing principle for unconventional superconductivity in two sublattice Kondo lattices.

\begin{acknowledgments}
We gratefully acknowledge discussions  David Vanderbilt, Kevin Lucht, Andreas Gleis,  Po-Jui Chen,  Liam Lau, Elena Hassinger, Joerg Schmalian,  Daniel Schultz and particularly, Philip Brydon. The early part of the work in this paper was completed under the support from the Division of Materials Research, US National Science Foundation grant DMR-1830707(PC). We acknowledge support from the US National Science Foundation (NSF) Grant Number 2201516 under the Accelnet program of Office of International Science and
Engineering (OISE)   
\end{acknowledgments}

\appendix
\counterwithin*{equation}{section} % reset 'equation' counter whenever '\section' is executed
\renewcommand\theequation{\thesection\arabic{equation}} % how to display the equation "number"

\section{Symplectic decoupling of the Heisenberg interaction}
\label{app:spn_decoupling}
Here we derive the dot product of the spins in the Heisenberg interaction. More specifically, the fermionic form of the Heisenberg interaction used in Eq.~\eqref{heisenberg_spn} will be obtained. The main framework for the symplectic $N$ construction comes from the work
Ref.~\cite{flintSymplecticTimeReversal2009}. For an even $N$, the spin can have $N$ number of components:
\begin{equation}
    \sigma=\pm1,\ldots,\pm N/2,
\end{equation}
and the antisymmetric tensor which time-reverses the spins is
\begin{equation}
    \epsilon^{\sigma}_{\ \sigma'}
    =\tilde{\sigma}\delta_{\sigma,-\sigma'}, \qquad
    \tilde{\sigma}\equiv {\rm sgn}(\sigma).
\end{equation}
There are $N(N+1)/2$ components of Hermitian spin generators $\mathcal S^a$ in  the Sp($N$) subgroup. These symplectic spin generators are odd under time reversal operation:
\begin{equation}
    \mathcal S^a=-\epsilon(\mathcal S^a)^T\epsilon^T,
\end{equation}
The normalization condition of the symplectic spin generators are defined as: 
\begin{equation}
    {\rm Tr}\left(\mathcal S^a\mathcal S^b\right)=2\delta^{ab}.
\end{equation}
The spin in terms of the fermionic operator and symplectic spin generator is given by:
\begin{equation}
    \hat S_i^a=f^\dagger_{i\alpha}\mathcal S^a_{\alpha\beta}f_{i\beta},
\end{equation}
In this convention, there is no additional factor of $1/2$.
For completeness, we now derive the Fierz identity used for decoupling the spin. An arbitrary $N\times N$ matrix $A$ can be decomposed into parts that are odd and even under time reversal as $A=A_S+A_A$, where
\begin{equation}
 A_S=-\hat\epsilon A_S^T\hat\epsilon^T,
 \qquad
 A_A=+\hat\epsilon A_A^T\hat\epsilon^T.
\end{equation}
The projector onto the symplectic subspace which has the property that inverts under time reversal is therefore
\begin{equation}\label{thedef}
 P_S A=\frac{1}{2}\left(A-\hat\epsilon A^T\hat\epsilon^T\right).
\end{equation}
Here $P_S$ represents the projector onto the symplectic subgroup.
Writing this relation in component form, 
\begin{equation}
    (P_S A)^{\alpha\beta}= P^{\alpha\beta}_{\gamma \delta} A^{\delta\gamma},
\end{equation}
then it follows from \eqref{thedef} that
\begin{equation}\label{projector_symplectic}
 (P_S)^{\alpha\beta}_{\gamma\delta}
 =\frac{1}{2}\left(
 \delta_{\alpha\delta}\delta_{\beta\gamma}
 -\epsilon_{\alpha \gamma}\epsilon_{\beta \delta}
 \right).
\end{equation}
The same projector $P_G$ can be expanded in terms of the symplectic part consisting of the complete set of Symplectic $N$ generators,
\begin{equation}
 A_S=\sum_a c_a\mathcal S^a.
\end{equation}
Using the normalisation $\mathrm{Tr}(\mathcal S^a\mathcal S^b)=2\delta_{ab}$ leads to
\begin{equation}
 c_a=\frac{1}{2}\mathrm{Tr}(A_S\mathcal S^a)
     =\frac{1}{2}\mathrm{Tr}(A\mathcal S^a),
\end{equation}
where the second equality comes from the fact that the symplectic and antisymplectic subgroups are orthogonal to each other. Then using the definition of the prefactors the projection onto the symplectic subgroup in terms of components:
\begin{align}
 (P_S A)^{\alpha\beta}
 &=\frac{1}{2}\sum_a
 \mathcal S^a_{\alpha\beta}\,
 \mathrm{Tr}(A\mathcal S^a)\\
 &=\frac{1}{2}\sum_a
 \mathcal S^a_{\alpha\beta}\mathcal S^a_{\gamma\delta}
 A^{\delta\gamma}.
\end{align}
Then the projector $P_G$ in terms of the spin generator representation has the form:
\begin{equation}\label{ap_eq:Ps_generator}
 (P_S)^{\alpha\beta}_{\gamma\delta}
 =\frac{1}{2}\sum_a
 \mathcal S^a_{\alpha\beta}\mathcal S^a_{\gamma\delta}.
\end{equation}
Combining both the definition of the projector in  Eq.~\eqref{projector_symplectic} and Eq.~\eqref{ap_eq:Ps_generator} leads to the symplectic completeness relation:
\begin{align}\label{fierz}
 \boxed{
 \sum_{a=1}^{N(N+1)/2}
 \mathcal S^a_{\alpha\beta}\mathcal S^a_{\gamma\delta}
 =
 \delta_{\alpha\delta}\delta_{\beta\gamma}
 -\epsilon_{\alpha \gamma}\epsilon_{\beta \delta}.}
\end{align}
This is  the Fierz identity for the SP$(N)$ generators. The second term imposes  the inversion of spins under the time reversal operation.

In the main text which uses the large-$N$ treatment, the Heisenberg interaction with the rescaled interaction  has the explicit form:
\begin{equation}
    H_H  =
    \frac{J_H}{N}
    \sum_{\langle ij\rangle}
    \sum_a
    \hat S_i^a\hat S_j^a .
\end{equation}
Using the completeness relation in Eq.~\eqref{fierz}, and normal ordering fermions on the two different sites leads to:
\begin{align}\label{app:heisenberg_fierz}
H_H=-\frac{J_H}{N}
\sum_{\langle ij\rangle, \sigma\sigma'
}&
\Big[
 f^\dagger_{j\sigma}f_{i\sigma}
 f^\dagger_{i\sigma'}f_{j\sigma'}+\tilde{\sigma}\tilde{\sigma}'
 f^\dagger_{i\sigma}f^\dagger_{j,-\sigma}
 f_{j,-\sigma'}f_{i\sigma'}
\Big],
\end{align}
where $\sigma,\sigma'=\pm1,\ldots,\pm N/2$ which are the spin components. This is the decoupled Heisenberg interaction in Eq.~\eqref{heisenberg_spn} of the main text.
Physically, the two terms in Eq.~\eqref{app:heisenberg_fierz} are the
particle hole and particle particle channels respectively. It is useful to introduce the bond orientation operators:
\begin{equation}
    D_{ji}=
    \sum_{\sigma}
    f^\dagger_{j\sigma}f_{i\sigma},
    \qquad
    P_{ji}  =
    \sum_{\sigma}
    \tilde{\sigma}\,
    f_{j,-\sigma}f_{i\sigma}.
\end{equation}
Since this is Hermitian construction, their Hermitian conjugates are given by:
\begin{equation}
    D^\dagger_{ji}
    =
    \sum_{\sigma}
    f^\dagger_{i\sigma}f_{j\sigma},
    \qquad
    P^\dagger_{ji}
    =
    \sum_{\sigma}
    \tilde{\sigma}\,
    f^\dagger_{i\sigma}f^\dagger_{j,-\sigma}.
\end{equation}
Rewriting the Heisenberg interaction in terms of the bond operators leads to the simple form:
\begin{equation}
\label{app:heisenberg_bonds}
    H_H=
    -\frac{J_H}{N}
    \sum_{\langle ij\rangle}
    \left(
        D^\dagger_{ij}D_{ij}
        +
        P^\dagger_{ij}P_{ij}
    \right).
\end{equation}
Equation~\eqref{app:heisenberg_bonds} gives the explicit form of the SU(2) isospin structure of the symplectic-$N$ representation. The antiferromagnetic exchange interaction generates degenerate particle hole and particle particle singlet decoupling channels. Using a Hubbard Stratonovich decoupling of the two channels leads to the $f$ electron hopping and pairing fields used in the main text.

\section{Charge conjugation transformation}
\label{app:charge_conjugation}
We now show explicitly develop the Balian-Werthammer (BW) isospin structure used in the main text.  The definitions of the fermionic operators used in the path integral formulation in the main text are:
\begin{equation}
    { C}_{{\bf k},\alpha,\sigma}= \pmat{
        c_{{\bf k},\alpha, \sigma}\cr c_{{\bf k},\alpha, -\sigma}\cr
        - c\dg_{-{\bf k},\alpha, 
        - \sigma}\cr c\dg_{-{\bf k},\alpha,  \sigma}}, \ 
    {F}_{{\bf k},\alpha,\sigma}=\pmat{
        f_{{\bf k},\alpha, \sigma}\cr f_{{\bf k},\alpha, -\sigma}\cr
        - f\dg_{-{\bf k},\alpha, 
        - \sigma}\cr f\dg_{-{\bf k},\alpha,  \sigma}},  
        \end{equation}
with $\sigma>0$. The Pauli matrices $\tau_i$ act in this particle hole also known as the isospin space. Combining the sublattice and the conduction and $f$ electron sectors leads to the spinor used in the mean field Hamiltonian,
\begin{equation}
 \Psi_{\bk}=\begin{pmatrix}
 C_{\bk +}\\ C_{\bk -}\\ { F}_{\bk +}\\ { F}_{\bk -}
 \end{pmatrix},
\end{equation}
To transform normal state  bilinears of the form $a^\dagger_{\bk}A(\bk)a_{\bk}$ into the Balian Werthammer form, we combine it with the corresponding hole component,  obtained by charge conjugation. The Balian Werthammer representation of this operator then becomes:
\begin{equation}
 A(\bk) \longrightarrow A(\bk)\frac12 (1+\tau_3)
 -\sigma_2 A^T(-\bk)\sigma_2 \frac12(1-\tau_3).
\end{equation}
Now in general, operators are either even, or odd under time-reversal, so that 
\begin{equation}
 \sigma_2 A^T(-\bk)\sigma_2=\left\{ \begin{array}{cc}
     + A(\bk)& \ \hbox{even}\cr
    - A(\bk)& \ \hbox{odd}\cr
 \end{array} \right . ,
 \label{eq:TR_Nambu_appendix}
\end{equation}
then the corresponding operators in the Balian-Werthammer  representation take the form 
\begin{equation}
    A(\bk) \rightarrow \left\{ \begin{array}{cc}
      A(\bk)\tau_3& \ \hbox{even}\cr
    A(\bk)& \ \hbox{odd}
 \end{array}\right.
\end{equation}
For example, in the normal state the Hamiltonian contains a dispersion and a Zeeman term ${\cal H}(\bk) = H_N(\bk) - \sigma_z B_z$. Since the first term is even under time-reversal, while the second is odd under time reversal, in the Balian Werthammer notation, 
\begin{equation}
    {\cal H}(\bk) \longrightarrow H_N(\bk)\tau_3 - \sigma_z B_z, 
\end{equation}
Thus, in the BW formulation, the  normal-state Hamiltonian takes the form 
\begin{equation}
    {\cal H}(\bk) = \pmat { 
    h_{\bk}\tau_3 -\sigma_z B& V \phi\dg(\bk)
    \cr V \phi(\bk)  & h^f_\bk\tau_3 -\sigma_z B }
\end{equation}
which accounts for the form used in \eqref{scham} of the main text. 

\bibliography{reference_new}

\end{document}